\documentclass[reprint,superscriptaddress,amsmath,amssymb,aps,pre]{revtex4-2}
\usepackage{graphicx}
\usepackage{dcolumn}
\usepackage{bm}
\usepackage{color}
\usepackage{units}
\usepackage{MnSymbol}
\usepackage{comment}
\usepackage{times}
\usepackage{soul}
\usepackage[utf8]{inputenc}
\usepackage{newunicodechar}
\usepackage{textcomp}
\usepackage[unicode=true,pdfusetitle,bookmarks=false,colorlinks=true,citecolor=blue,urlcolor=blue,linkcolor=blue]{hyperref}
\begin{document}

\title{Critical Motility-Induced Phase Separation in Three Dimensions is Consistent with Ising Universality}

\author{Jiechao Feng}
\affiliation{Graduate Group in Applied Science \& Technology, University of California, Berkeley, California 94720, USA}

\author{Daniel Evans}
\affiliation{Department of Materials Science and Engineering, University of California, Berkeley, California 94720, USA}

\author{Ahmad K. Omar}
\email{aomar@berkeley.edu}
\affiliation{Department of Materials Science and Engineering, University of California, Berkeley, California 94720, USA}
\affiliation{Materials Sciences Division, Lawrence Berkeley National Laboratory, Berkeley, California 94720, USA}

\begin{abstract}
Identifying the universality class of critical active phase transitions has been the subject of recent interest and controversy.
Resolving these controversies will require robust numerical investigations to determine whether active critical exponents point to novel universality classes or are consistent with established ones. 
Here, we conduct large-scale computer simulations and a finite-size scaling analysis of the motility-induced phase separation (MIPS) of active Brownian hard spheres in three dimensions (3D), finding that the static and dynamic critical exponents all closely match those of the 3D Ising universality class with a conserved scalar order parameter.
This finding is corroborated by a fluctuating hydrodynamic description of the critical dynamics of the order parameter field which flows to the Wilson-Fisher fixed point in three dimensions.
Our work suggests that 3D MIPS and likely the entire phase diagram of active Brownian hard spheres is similar to that of molecular passive fluids despite the absence of Boltzmann statistics.
\end{abstract}

\maketitle
\section{Introduction}
In thermodynamic equilibrium, critical points of continuous phase transitions have been understood for over a half century~\cite{Kadanoff67,Fisher64,Hohenberg77,Kardar07,Tauber14}.
Despite what can appear to be vastly disparate microscopic details, systems with the same symmetries and satisfying the same conservation laws will behave remarkably similar near their respective critical points~\cite{Kardar07,Stanley99,Tauber14}.
This behavior includes how the correlation length and response functions diverge (governed by \textit{static} critical exponents) and even how the dynamics slow down (set by \textit{dynamic} exponents~\cite{Odor04}) as the critical point is approached.
The universality of critical phenomena is not restricted to thermodynamic equilibrium. 
The study of critical phenomena in intrinsically nonequilibrium systems has revealed novel fixed points that belong to universality classes exclusive to driven~\cite{Katz1983PhaseSystems,Kardar1986DynamicInterfaces,Cardy1996TheoryWalks,Dornic2001CriticalModel,Takeuchi07} and/or active~\cite{Toner95,Toner98,Toner12,Alert20,Cagnetta22,Jentsch24,Miller24} systems.

The observation of activity-induced phase transitions in recent years has become routine, with many of the transitions exhibiting a critical point. 
Among the more well-studied active transitions, known as ``motility-induced phase separation'' (MIPS), is the phenomena in which nonaligning and purely repulsive active particles may phase separate into coexisting domains of disordered fluids~\cite{Tailleur2008,Fily12,Redner13,Buttinoni13,Cates15}. 
The similarity of MIPS to a traditional liquid-gas transition (i.e.,~bulk coexistence solely characterized by a conserved scalar order parameter) is suggestive of Ising universality.
Several investigations have sought to numerically determine the universality class of MIPS in two dimensions (2D) using both on-lattice~\cite{Partridge19,Dittrich21} and off-lattice~\cite{Siebert18,Maggi21} simulations. 
While the results of some of these studies are indeed consistent with Ising universality~\cite{Partridge19,Maggi21}, others have been less conclusive~\cite{Siebert18,Dittrich21}. 
Caballero~\textit{et al.}~\cite{Caballero18} recently used a one-loop dynamical renormalization group (RG) approach to show that active model B+ (AMB+)~\cite{Nardini17,Tjhung18}, a nonequilibrium scalar field theory, can exhibit a novel fixed point that may be associated with microphase separation in addition to the Wilson-Fisher fixed point.
The coalescence of these fixed points in two dimensions can alter the universality class and has been suggested as a possible origin of these conflicting numerical investigations~\cite{Speck2022CriticalTheories}. 

In this article, we aim to shed light on the nature of the MIPS critical point by providing a large-scale numerical investigation of MIPS in \textit{three dimensions}~\cite{Wysocki2014, Stenhammer14, VanDamme2019, Nie2020, Omar20, Turci2021, Omar21} in systems of active Brownian hard spheres. 
Our extensive finite-size scaling analysis results in critical exponents ($\nu$, $\gamma$, $\beta$, and $z$) consistent with the 3D Ising universality class with a conserved scalar order parameter. 
We develop a fluctuating hydrodynamic description of critical MIPS which, after mapping to AMB+, can be shown to flow to the Wilson-Fisher fixed point in 3D by following the approach of Caballero~\textit{et al.}~\cite{Caballero18}.
This suggests that both the static and dynamic critical exponents of 3D MIPS in systems of hard active Brownian particles (ABPs) should indeed belong to the Ising universality class.
Our work provides additional evidence that the 3D phase diagram of ABPs is likely similar to that of simple substances in thermodynamic equilibrium.

\begin{figure*}
	\centering
	\includegraphics[width=.95\textwidth]{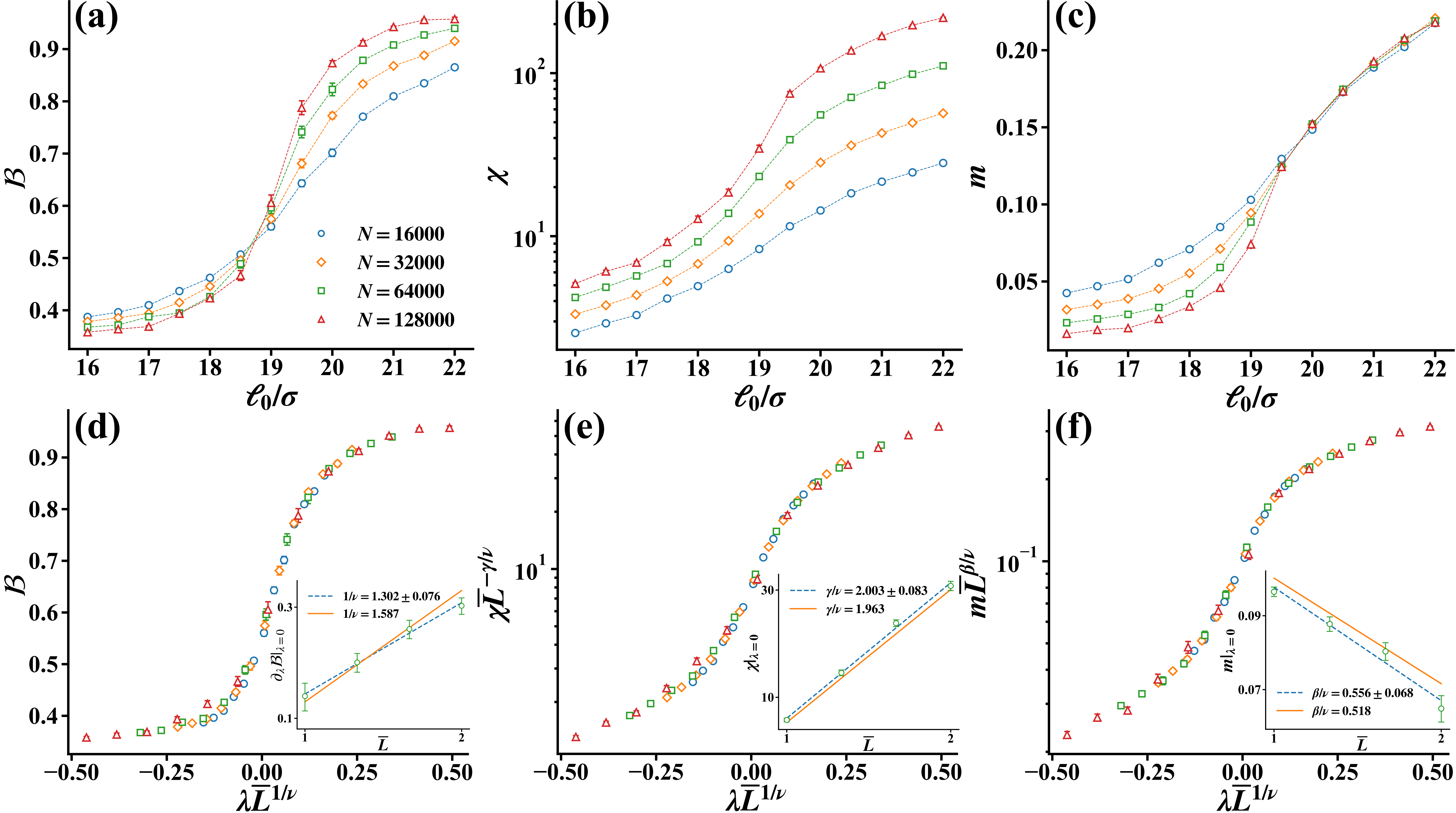}
	\caption{\protect\small{{Finite-size scaling analysis for determining static critical exponents. System size and activity dependence of (a) $\mathcal{B}$, (b) $\chi$, and (c) $m$. (d) Collapse of data in (a) as a function of the scaling variable $\lambda \overline{L}^{1/\nu}$ with $\nu=0.630$ [Inset: Slope of $\mathcal{B}$ at $\lambda=0$ ($\ell_0=\ell_{\rm c}$) as a function of system size, where the dashed line is the best fit of $1/\nu=1.302(0.076)$ and the solid line is $1/\nu=1.587$]. (e) Collapse of data in panel (b) with the exponents $\gamma=1.237$ and $\nu=0.630$ [Inset: $\chi$ at $\lambda=0$ as a function of system size, where the dashed line is the best fit of $\gamma/\nu=2.003(0.083)$ and the solid line is $\gamma/\nu=1.963$]. (f) Collapse of data in panel (c) with the exponents $\beta=0.326$ and $\nu=0.630$ [Inset: $m$ at $\lambda=0$ as a function of system size, where the dashed line is the best fit of $\beta/\nu=0.556(0.068)$ and the solid line is $\beta/\nu=0.518$]. We define $\overline{L}=L/L_0$ where $L_0/\sigma \approx 9.2$ is the size of the smallest sub-box ($N=1.6 \times 10^4$).}}}
	\label{Fig:1} 
\end{figure*}

\section{Simulation methods}
We consider the simplest model that captures the critical MIPS behavior: 3D athermal active Brownian hard spheres~\cite{Omar21,Feng2024TheoryMatter}.
Each of the $N$ particles experiences three translational forces: a drag force $-\zeta_{\rm d} \dot{\mathbf{x}}$ proportional to the particle velocity $\dot{\mathbf{x}}$ and translational drag coefficient $\zeta_{\rm d}$; a conservative interparticle force $\mathbf{F}^{\rm C}[\mathbf{x}^N]$ where $\mathbf{x}^N$ represents all particle positions; and an active force $\mathbf{F}^{\rm A}=\zeta_{\rm d} U_0 \mathbf{q}$, where $U_0$ is the intrinsic active speed.
Stochastic Brownian translational forces, which simply attenuate the influence of activity, are neglected. 
The particle orientation $\mathbf{q}$ independently follows diffusive dynamics $\dot{\mathbf{q}}=\bm{\Omega}\times \mathbf{q}$ where the stochastic angular velocity has mean $\bm{0}$ and variance $\langle \bm{\Omega}(t)\bm{\Omega}(0) \rangle = 2/\tau_{\rm R} \delta(t) \mathbf{{I}}$, where $\tau_{\rm R}$ represents the characteristic reorientation time (or inverse rotational diffusion).
The interparticle force $\mathbf{F}^{\rm C}[\mathbf{x}_N; \sigma, \varepsilon]$ is taken from a Weeks-Chandler-Anderson potential, which is characterized by a Lennard-Jones diameter $\sigma$ and energy $\varepsilon$~\cite{Weeks71}.
Taking $\zeta_{\rm d} U_0$, $\sigma$, and $\tau_{\rm R}$ to be the characteristic units of force, length, and time, respectively, the dimensionless Langevin equation follows as
\begin{equation}
    \label{eq:eom}
     \overline{\dot{\mathbf{x}}} = \frac{\ell_0}{\sigma} \left ( \mathbf{q} + \overline{\mathbf{F}}^{\rm C} [ \overline{\mathbf{x}}^{\rm N}; \mathcal{S}] \right ),
\end{equation}
where we have defined the intrinsic ``run length'' (activity) $\ell_0 \equiv U_0 \tau_{\rm R}$.
The dimensionless force $\overline{\mathbf{F}}^{\rm C}$ depends on the reduced positions $\overline{\mathbf{x}}^{\rm N}$ and is fully characterized by the ``stiffness'' parameter $\mathcal{S} \equiv \varepsilon/(\zeta_{\rm d} U_0 \sigma)$.
We use $\mathcal{S}=50$ to effectively achieve hard-sphere statistics.
The particles behave precisely as hard spheres in this limit with diameter $d_{\rm hs}=2^{1/6}\sigma$, despite the use of a continuous potential. 
With $\mathcal{S}$ fixed, the system state is fully described by the volume fraction $\phi\equiv N v_D/V$ and activity $\ell_0/\sigma$, where $v_D\equiv \pi d_{\rm hs}^3/6$ is the volume of a single particle.

To determine the static critical exponents, we perform simulations with fully periodic boundary conditions using asymmetric boxes with an aspect ratio of 1:1:4 ($L_z=4L_x=4L_y$).
When above the critical activity, this choice of simulation geometry is found to result in an interface with an average normal direction along the $z$ dimension (consistent with area-minimizing interfaces~\cite{Langford24}).
We consider systems of sizes $N=(1.6,3.2,6.4,12.8)\times 10^4$ and examine the statistics of volume fraction fluctuations within subsystems~\cite{Binder81,Binder1987FiniteTransitions,Rovere88,Rovere90,Rovere93}. 
To avoid spurious effects due to the presence of an interface~\cite{Siebert18,Partridge19,Dittrich21,Maggi21}, we compute the quantities of interest only in four cubic sub-boxes of size $L=L_x/2$ in each of the centers of the dense (liquid) and dilute (gas) phases, allowing us to avoid the interface.
All simulations are performed at a fixed overall volume fraction of $\phi=0.48$, the estimated critical density~\cite{Omar21, Omar23b}. 
All simulations were conducted using \texttt{HOOMD-blue}~\cite{Anderson2020HOOMD-blue:Simulations}.
Additional simulation details can be found in the Supplemental Material (SM)~\footnote{See Supplemental Material at [URL], which includes Refs.~\cite{Brezin74,Family90,Roman1997FluctuationsEffects,Lopez97,Lassig98,Roman98,Landau2013Statistical5,Virtanen2020SciPyPython}, for theory of critical fluctuating hydrodynamics for ABPs, numerical details of finite-size scaling analysis, and numerical determination of critical exponents}.

\section{Results}
\subsection{Static critical exponents} 
As the critical activity $\ell_c$ is approached, the correlation length associated with the order parameter diverges as $\xi \sim \xi_{\pm} |\lambda|^{-\nu}$ where we have defined the reduced activity $\lambda\equiv (\ell_0-\ell_{\rm c})/\ell_c$. 
The $\pm$ subscript denotes a distinction in properties and functions depending on if $\lambda$ is greater than or less than 0.
We consider observables $\mathcal{O}$ that are expected to diverge (or vanish) at the critical point in the absence of finite-size effects.
The finite-size scaling ansatz posits the form of these observables, to leading order, as $\mathcal{O}=|\lambda|^{-\zeta_{\mathcal{O}}} F_{\mathcal{O} \pm}(\xi/L)$, where $\zeta_{\mathcal{O}}$ and $F_{\mathcal{O}\pm}$ are the associated critical exponent and scaling functions.
Finite system sizes prevent these observables from truly diverging at the critical point and, for $\xi \gg L$, we thus expect $\mathcal{O}=L^{\zeta_{\mathcal{O}}/\nu} \Tilde{F}_{\mathcal{O} \pm}(|\lambda|(L/\xi_{\pm})^{1/\nu})$~\cite{Tauber14}.
This implies that all values of $\mathcal{O}$ measured for different system sizes should collapse when $\mathcal{O} L^{-\zeta_{\mathcal{O}}/\nu}$ is plotted as a function of $\lambda L^{1/\nu}$ and should scale as $L^{\zeta_{\mathcal{O}}/\nu}$ precisely at $\ell_0=\ell_c$.
We consider three observables which are related to the statistics of volume fraction fluctuations within sub-boxes. 
The first is the fourth-order (Binder) cumulant defined as $\mathcal{B} \equiv \langle \Delta \phi^2\rangle^2_L/\langle \Delta \phi^4\rangle_L$, where $\Delta \phi =\phi-\langle \phi \rangle_L$, $\langle \phi \rangle_L$ is the average volume fraction of all sub-boxes, and $\langle ... \rangle_L$ indicates an average over configurations and all sub-boxes.
The second is the susceptibility, $\chi \equiv (L^3/v_D) \langle \Delta \phi^2\rangle_L/\langle \phi \rangle_L$, which is proportional to the variance of volume fraction fluctuations (again, the expectation is using the distribution computed over all sub-boxes).
The final observable we consider is the difference between the average volume fraction of the two phases defined as $m\equiv \langle \phi \rangle_{L, {\rm l}}-\langle \phi \rangle_{L, {\rm g}}$, where $\langle ... \rangle_{L,{\rm l}}$ and $\langle ... \rangle_{L, {\rm g}}$ denote the average over liquid and gas sub-boxes, respectively.
The scaling exponents (i.e.,~$\zeta_{\mathcal{O}}$) for our three observables of $\mathcal{B}$, $\chi$, and $m$ are $0$, $\gamma$, and $-\beta$, respectively. 

The Binder cumulant (straightforwardly related to the Binder parameter~\cite{Binder81}) allows us to identify the location of the critical point as $\mathcal{B}$ is size independent at $\ell_0 = \ell_c$.
Figure~\ref{Fig:1}(a) displays the activity dependence of $\mathcal{B}$ for several system sizes with all curves intersecting between $18.5<\ell_c/\sigma<19.0$, an uncertainty that is appreciably smaller than in previous 2D simulations~\cite{Siebert18, Maggi21}.
The critical activity is estimated to be  $\ell_c/\sigma = 18.85 (0.05)$ using the two largest system sizes. 
Using this value and the 3D Ising exponent $\nu=0.630$~\cite{Hasenbusch1999CriticalActions}, all $\mathcal{B}$ curves nicely collapse onto a universal curve [see Fig.~\ref{Fig:1}(d)].
We also extract the exponent $\nu$ from the derivative of $\mathcal{B}$ at criticality: $d\mathcal{B}/d\lambda|_{\lambda=0}\propto L^{1/\nu}$.
Importantly, differentiating $\mathcal{B}$ near the critical point likely leads to considerable error. 
Indeed, we find $\nu=0.768(0.045)$, a marked deviation from the value for Ising universality.
Our measured discrepancy of $\nu$ between simulations and the Ising value is smaller than that measured in Siebert~\textit{et al.}'s 2D simulations~\cite{Siebert18} but larger than that reported by Maggi~\textit{et al.}~\cite{Maggi21}. 
Below, we offer an alternative measure for determining $\nu$ that avoids differentiation and results in an exponent that is much more consistent with Ising universality.

We next study the scaling of susceptibility $\chi$.
The raw data and the scaled results of $\chi$ are shown respectively in Figs.~\ref{Fig:1}(b) and (e).
We again find that 3D Ising exponents (now introducing $\gamma=1.237$~\cite{Hasenbusch1999CriticalActions})  lead to good collapse.
At criticality, we expect $\chi|_{\lambda=0}\propto L^{\gamma/\nu}$.
Using this relation, we extract $\gamma/\nu=2.003(0.083)$, which is in excellent agreement with the 3D Ising value and well within our statistical uncertainty.
We note that in previous simulation studies of MIPS critical scaling, this is generally the exponent with the smallest discrepancy from the Ising value~\cite{Siebert18,Partridge19,Dittrich21,Maggi21}.

Finally, we determine $\beta$ from the scaling of the phase volume fraction difference $m$.
Figure~\ref{Fig:1}(c) shows the raw data, where we see immediately that $m$ becomes size independent above the critical point.
Figure~\ref{Fig:1}(f) again shows good data collapse using the 3D Ising exponents including $\beta=0.326$~\cite{Hasenbusch1999CriticalActions}.
At criticality, we expect $m\propto L^{-\beta/\nu}$, from which we extract $\beta/\nu =0.556(0.068)$, which is close to the Ising value.
We can also obtain the exponent $\beta$ from $m\propto \lambda^{\beta}$ for $\lambda \gtrsim 0$, finding $\beta=0.347(0.008)$ for the largest system, which is again relatively close to the Ising value~\cite{Note1}.
From our extracted value of $\beta/\nu$ and $\beta$, we estimate $\nu=0.624(0.078)$, which is nearly exactly the Ising value (in contrast to our earlier estimate using $d\mathcal{B}/d\lambda|_{\lambda=0}$).
We note that for 2D MIPS, $\beta$ appears to be the most controversial exponent: Ref.~\cite{Siebert18} found $\beta$ is appreciably larger than the 2D Ising value, while the on-lattice simulation in Ref.~\cite{Dittrich21} also showed large discrepancies of both $\beta$ and $\beta/\nu$ with Ising values.
The relatively small discrepancies between all of our measured exponents and Ising exponents in comparison to 2D studies may point towards a genuine distinction between 2D and 3D MIPS.

\begin{figure}
	\centering
    \includegraphics[width=.45\textwidth]{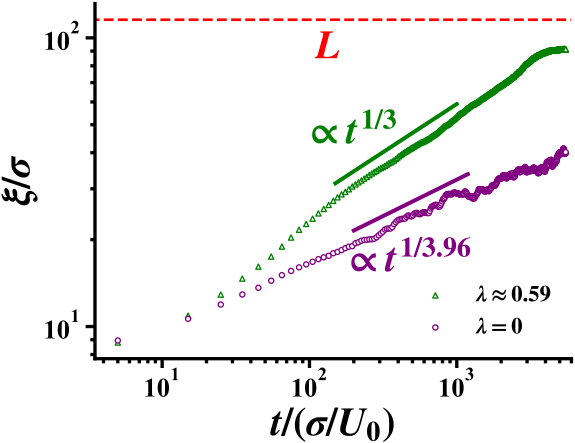}
	\caption{\protect\small{{The correlation length $\xi(t)$ at criticality ($\lambda=0$, purple circles) and inside phase-separation regime ($\lambda\approx 0.59$, green triangles), as determined from the inverse of the first moment of the static structure factor. Each point represents the average of ten consecutive data points, computed at intervals of $\sigma/U_0$. The purple line shows $t^{1/z}$, where $z=4-\eta\approx 3.964$ for model B in 3D. The green line shows $t^{1/3}$ which is the classical scaling for Ostwald ripening and/or coalescence. The red dashed line denotes the upper limit for $\xi$, which is the system size $L$.}}}
	\label{Fig:2} 
\end{figure}

\subsection{Dynamic critical exponent} 
While the static exponents of 3D MIPS are consistent with Ising universality, systems sharing the same static critical exponents can exhibit different dynamic exponents, depending on their equations of motion.
This raises the question of whether activity could manifest through a modification of the critical dynamics. 
Approaching the critical point, as the correlation length diverges, one expects the associated dynamical relaxation time to grow as well with $\tau \sim \xi^z$ where we have defined the dynamic critical exponent $z$~\cite{Hohenberg77,Tauber14}.
Here, $\tau$ is the relaxation time of spatial correlations of the order parameter on length scales of $\xi$. 
For finite system sizes, the time evolution of the correlation length as it approaches its steady state value offers insight into this dynamic exponent. 
Dimensional analysis suggests the following $\xi(t) = \xi(t\rightarrow \infty)\mathcal{F}_t(t/\tau)$ where $\mathcal{F}_t$ is a scaling function and we have assumed an initial condition of a random quench with a small correlation length. 
This scaling hypothesis is thus only anticipated to hold beyond some initial duration after the quench, $\tau_0$.
Near the critical point, the finite system size bounds the terminal correlation length with $\xi(t\to \infty) = L$ and thus also bounds the relaxation time to $\tau \sim L^z$.
Prior to reaching this terminal value the correlation length follows $\xi(t) = L\mathcal{F}_t(t/\tau)$.
For $t \ll \tau$ (while $t > \tau_0$) the time-dependent correlation length should be independent of the system size. 
In this regime, a power-law expansion of the scaling function must take the form $\mathcal{F}_t(t/\tau) \sim (t/\tau)^{1/z}$ to ensure the independence of $\xi(t)$ with the system size.
We therefore anticipate a window of time ($\tau_0 < t \ll \tau$) in which $\xi(t) \sim t^{1/z}$. 
In the case of model B dynamics, the longest relaxation time is set by the diffusive motion with $\tau \sim \xi^2/D$. 
This diffusivity is the collective (or chemical) diffusivity that, in the case of a single conserved scalar order parameter, can be expressed as the product of an Onsager transport coefficient, $M$, and the inverse susceptibility with $D \sim M/\chi$.
If we assume that the Onsager mobility displays no singular behavior as the critical point is approached, we then have $\tau \sim \xi^2\chi \sim \xi^{2+\gamma/\nu}$ and the critical exponent is entirely determined from the static exponents with $z = 2+\gamma/\nu \equiv 4 - \eta$ ($z \approx 3.964$ or $\eta \approx 0.036$ for 3D Ising universality with conserved dynamics).
We note that previous studies on 2D MIPS have found the critical dynamics of MIPS belong to the 2D Ising universality class with conserved dynamics ($\eta=1/4$~\cite{Partridge19,Dittrich2023GrowthSystems}), but our work focuses on determining $z$ for critical 3D MIPS.

To numerically determine $z$ for critical MIPS, we first need to define the correlation length from simulations.
We follow previous works~\cite{Stenhammar2013ContinuumParticles,Stenhammer14,Wittkowski14,Mandal19} and determine the correlation length $\xi(t)$ from the inverse of the first moment of the static structure factor~\cite{Note1}.
Figure~\ref{Fig:2} presents the time dependence of this length scale after a quench to the critical point ($\lambda = 0$) for our largest system size $N=10^6$.
While no power-law scaling is observed at early or late times (as anticipated), we do find an intermediate time window in which the correlation length scales with time in agreement with 3D Ising universality under model B dynamics, characterized by $z \approx 3.96$ (see SM for detailed analysis~\cite{Note1}). 
This behavior is contrasted with a deep quench ($\lambda \approx 0.59$ in Fig.~\ref{Fig:2}) which reveals a scaling exponent of $1/3$, consistent with classical theories of Ostwald ripening and/or coalescence~\cite{Chaikin1995PrinciplesPhysics,Bray2002TheoryKinetics,Puri2004KineticsTransitions,Stenhammer14}.
It thus appears that both the static and dynamic exponents of critical 3D MIPS in systems of ABPs are entirely consistent with those of the Ising universality class.

\subsection{Coarse-grained dynamics} 
The consistency of our numerical results with 3D Ising universality motivates a theoretical examination of the fluctuating hydrodynamic equations of active Brownian spheres in order to carefully analyze the fixed points. 
The dynamics of the order parameter can be obtained exactly from systematic coarse-graining.
These exact dynamics for a specific field are, for all but the simplest cases, coupled to additional fields. 
However, in certain limits, it may be possible to formally approximate these dynamics through closures that reduce the number of fields. 
The resulting dynamics can perhaps be connected to phenomenological dynamical descriptions. 
Here, we provide an analysis of the near-critical coarse-grained order-parameter dynamics and, after mapping the dynamics to AMB+~\cite{Tjhung18, Caballero18}, show that critical MIPS in systems of 3D ABPs falls under the Ising universality class (we provide more details in the SM~\cite{Note1}).

We recently derived an \textit{approximate} fluctuating hydrodynamic description of overdamped ABPs~\cite{Langford24} under the conditions of small fluxes, small spatial gradients, and long times.
We expect that near-critical dynamics meet these conditions. 
Within these limits, the dynamics for the density field can be expressed solely in terms of the local density:
\begin{subequations}
\label{eq:fluc_hydro}
\begin{align}
    \frac{\partial \rho}{\partial t} &= - \bm{\nabla} \cdot \mathbf{J},  \\
    \mathbf{J} &= \frac{1}{\zeta_{\rm d}} \bm{\nabla} \cdot \left( \bm{\sigma}^{\rm C} + \bm{\sigma}^{\rm act} \right) + \bm{\eta}^{\rm act}, \\
    \bm{\sigma}^{\rm C} &= \left[ -p_{\rm C} (\rho)+\kappa_1(\rho) \nabla^2 \rho + \kappa_2(\rho) |\nabla \rho|^2 \right] \mathbf{I}  \nonumber \\
    &+\kappa_3(\rho)\bm{\nabla} \rho \bm{\nabla} \rho + \kappa_4(\rho) \bm{\nabla} \bm{\nabla} \rho , \label{eq:sigma_c} \\
    \bm{\sigma}^{\rm act} &= \left[ - p_{\rm act} (\rho) + a(\rho) \nabla^2 \rho \right] \mathbf{I} +b(\rho) \bm{\nabla} \rho \bm{\nabla} \rho,\label{eq:sigma_act} \\
    \langle \bm{\eta}^{\rm act} (\mathbf{r}, t) & \bm{\eta}^{\rm act} (\mathbf{r}', t') \rangle = 2 \frac{k_B T^{\rm act}}{\zeta_{\rm d}} \left( \rho \mathbf{I} - \frac{d}{d-1} \mathbf{Q}' \right) \nonumber\\
    & \quad \qquad \qquad \times \delta(t-t') \delta(\mathbf{r} - \mathbf{r}'), \label{eq:noise} \\
    \mathbf{Q}' &= \frac{\rho U_0}{d p_{\rm act}} \left( -a(\rho)  \nabla^2 \rho \mathbf{I} - b(\rho) \bm{\nabla}\rho \bm{\nabla} \rho \right) .\label{eq:traceless_nematic}
\end{align}
\end{subequations}
In Eq.~\eqref{eq:fluc_hydro}, $\rho(\mathbf{r}, t)$ is the coarse-grained density field, $\mathbf{J}$ the density flux, $\zeta_{\rm d}$ the single-particle translational drag coefficient, $\bm{\sigma}^{\rm C}$ the interparticle conservative stress, $\bm{\sigma}^{\rm act}$ the active effective stress, $\bm{\eta}^{\rm act}$ the athermal stochastic contribution to density flux, $\mathbf{Q}'$ the traceless nematic order density, and $d$ the number of spatial dimensions.
The anisotropic noise presented above is entirely athermal with the amplitude set by the active energy scale $k_B T^{\rm act} \equiv \zeta_{\rm d} U_0 \ell_0 /d(d-1)$.
The form of the noise presented in Eq.~\eqref{eq:noise} is exact for the \textit{microscopic} density and nematic field and approximate for the \textit{coarse-grained} fields of interest.
$\bm{\sigma}^{\rm C}$ [Eq.~\eqref{eq:sigma_c}] and $\bm{\sigma}^{\rm act}$ [Eq.~\eqref{eq:sigma_act}] are expanded into bulk terms ($p_{\rm C}$ and $p_{\rm act}$ respectively) and density gradients up to the second order, with $\{ \kappa_i (\rho) \}$, $a(\rho)$, and $b(\rho)$ expansion coefficients.
As Ref.~\cite{Omar23b} considered \textit{repulsive} ABPs at high activities, the interfacial coefficients associated with the interaction stress ($\{ \kappa_i(\rho) \}$) were discarded as they fundamentally scale with the particle size ($\sigma$) while those associated with the active stress [$a(\rho)$ and $b(\rho)$] scale with the run length ($\ell_0$).
The forms of $\{ \kappa_i(\rho) \}$ can be found for pairwise interacting systems by formally expanding the spatial kernel of the interaction stress (see, for example, Ref.~\cite{Chiu2024TheoryCoexistence}).

We now look to express our fluctuating hydrodynamics model in a form that facilitates a comparison with AMB+.
Before expanding our dynamics near the critical point, we introduce the order parameter ${\varphi(\mathbf{r}, t) \equiv \left(\rho(\mathbf{r}, t)-\rho_{\rm c} \right)v_D}$, which has the (lack of) dimensions of a volume fraction, where $\rho_{\rm c}$ is the critical density.
Defining the dynamic pressure ${\mathcal{P}(\rho)=p_{\rm C}(\rho)+p_{\rm act}(\rho)}$, the critical point can be determined from the following equations~\cite{Omar23b}:
\begin{equation}
\label{eq:Critical}
    \frac{\partial \mathcal{P}}{\partial \rho}\Bigg|_{\rho_{\rm c}, \ell_{\rm c}}=\frac{\partial^2 \mathcal{P}}{\partial \rho^2}\Bigg|_{\rho_{\rm c}, \ell_{\rm c}}=0 .
\end{equation}
As the system approaches the critical point, Eq.~\eqref{eq:fluc_hydro} can be expanded around the critical density.
This expansion yields the dynamics for the order parameter $\varphi(\mathbf{r}, t)$.
For Taylor expanding $\mathcal{P}(\rho)$ near $\rho_{\rm c}$, we have
\begin{align}
\label{eq:P_expansion}
    \mathcal{P}(\rho) &=\mathcal{P}_{\rm c}+ g_1 \varphi(\mathbf{r}, t)+g_2 \varphi^2(\mathbf{r}, t) +g_3 \varphi^3(\mathbf{r}, t)+ ... \ ,
\end{align}
where $g_n =(v_D^n n!)^{-1} \partial^n \mathcal{P} / \partial \rho^n$ with the derivatives evaluated at the critical point.
From Eq.~\eqref{eq:Critical} we have $g_1=g_2=0$; however we retain $g_1$ for clarity as the term it corresponds to is present in AMB+ (albeit 0 at the critical point).
Furthermore, we Taylor expand $a = a_0 + a_1 \varphi$ where $a_0$ and $a_1 \equiv \partial_{\rho} a / v_D$ are $a$ and its derivative with respect to $\varphi$, both evaluated at the critical point.
Substituting these expansions for $\mathcal{P}$ and $a$ into Eq.~\eqref{eq:fluc_hydro} and truncating the dynamics at $\mathcal{O}(\nabla^4 \rho^2)$ [along with a single cubic term of $\mathcal{O}(\nabla^2 \rho^3)$], we find
\begin{multline}
    \label{eq:varphiEOM}
    \frac{\partial \varphi}{\partial t}= \frac{1}{\zeta_d} \boldsymbol{\nabla} \cdot \bigg( v_D \left( g_1 + 3 g_3 \varphi^2 \right) \boldsymbol{\nabla} \varphi - \left( a_0 + a_1 \varphi \right) \boldsymbol{\nabla} \left(\nabla^2 \varphi \right) \\ - \left( a_1 + \frac{b_0}{v_D}  \right) \nabla^2 \varphi \boldsymbol{\nabla} \varphi  -  \frac{b_0}{v_D}\left(\nabla \varphi \cdot \nabla \right) \boldsymbol{\nabla} \varphi + \zeta_d \boldsymbol{\eta}^{\rm act} \bigg),
\end{multline}
where $b_0$ is the value of $b$ at the critical point.

To analyze the critical fluctuations, one might first perform a standard dimensional analysis of the coarse-grained dynamics [e.g., Eq.~\eqref{eq:varphiEOM}] about a fixed point and discard irrelevant terms.
It is typically then assumed that the remaining terms govern the critical dynamics and thus enable a straightforward identification of the characteristic phenomenological model—and hence the universality class—of the critical point.
However, such an approach may break down under certain circumstances.
For example, nonlinear terms which are formally irrelevant might induce a new strong-coupling fixed point when their value is large enough, with the Kardar-Parisi-Zhang equation in $d>2$ serving as a well-known example~\cite{Kardar1986DynamicInterfaces,Tauber14,Canet2010NonperturbativeEquation}.
A central message of Ref.~\cite{Caballero18} is that, through a careful one-loop RG analysis, the active terms in AMB+ which are irrelevant near the Wilson-Fisher fixed point could induce a new fixed point and a strong-coupling regime where the nonlinear parameters diverge.

Before mapping to AMB+, we identify the natural units of time and length as $\tau_{\rm R}$ and $\ell_c$ and subsequently nondimensionalize Eq.~\eqref{eq:varphiEOM} with this time and length scale, leading to the following equation of motion:
\begin{multline}
    \label{eq:varphiEOM2}
    \frac{\partial \varphi}{\partial \overline{t}}= \frac{1}{\zeta_d} \overline{\boldsymbol{\nabla}} \cdot \bigg( \left[ \overline{g}_1 + 3 \overline{g}_3 \varphi^2 \right] \overline{\boldsymbol{\nabla}} \varphi - \left( \overline{a}_0 + \overline{a}_1 \varphi \right) \overline{\boldsymbol{\nabla}} \left(\overline{\nabla}^2 \varphi \right) \\ - \left( \overline{a}_1 + \overline{b}_0   \right) \overline{\nabla}^2 \varphi \overline{\boldsymbol{\nabla}} \varphi  -  \overline{b}_0 \left(\overline{\nabla} \varphi \cdot \overline{\nabla}\right) \overline{\boldsymbol{\nabla}} \varphi - \zeta_d \boldsymbol{\eta}^{\rm act} \bigg),
\end{multline}
where $\overline{t}\equiv t / \tau_{\rm R}$, $\overline{\nabla}\equiv \ell_{\rm c}\nabla$, $\overline{g}_n\equiv v_D g_n/U_0\ell_{\rm c}$, $\overline{a}_n\equiv a_n/U_0\ell_{\rm c}^3$, and $\overline{b}_0\equiv b_0/v_D U_0\ell_{\rm c}^3$.
The form of AMB+ used in Ref.~\cite{Caballero18} is
\begin{multline}
    \label{eq:AMB+eq}
    \frac{\partial \varphi}{\partial t} = \boldsymbol{\nabla} \cdot \bigg(M \bigg[ \boldsymbol{\nabla}\left( r \varphi + u \varphi^3 - K \nabla^2 \varphi + \lambda | \nabla \varphi|^2 + \frac{\nu}{2} \nabla^2(\varphi^2)\right) \\ + \zeta \nabla^2 \varphi \boldsymbol{\nabla} \varphi \bigg] - \sqrt{2 D M} \boldsymbol{\Lambda} \bigg) \\ = \boldsymbol{\nabla} \cdot \bigg(M \bigg[ \left( r + 3 u \varphi^2 \right) \boldsymbol{\nabla} \varphi - \left(K - \nu \varphi \right) \boldsymbol{\nabla} \left( \nabla^2 \varphi \right) - \left( \zeta - \nu \right) \nabla^2 \varphi \boldsymbol{\nabla} \varphi \\ + 2 \left( \lambda + \nu \right)  \left(\nabla \varphi \cdot \nabla \right) \boldsymbol{\nabla} \varphi \bigg] - \sqrt{2 D M} \boldsymbol{\Lambda} \bigg),
\end{multline}
where $r$, $u$, $K$, $\nu$, $\lambda$, $\zeta$, $M$, and $D$ are all constants and $\boldsymbol{\Lambda}$ is isotropic white noise with zero mean and unit variance.
It is clear that the deterministic part of the dynamics in Eq.~\eqref{eq:varphiEOM2} can be mapped to AMB+ [Eq.~\eqref{eq:AMB+eq}], setting $M=1$ as was done in Ref.~\cite{Caballero18}, with
\begin{subequations}
\label{eq:mapping}
\begin{align}
r &= \frac{\overline{g}_1}{\zeta_{\rm d}}, \\
u &= \frac{\overline{g}_3}{\zeta_{\rm d}}, \\
K &= \frac{\overline{a}_0 }{\zeta_{\rm d}}, \\
\nu &= -\frac{\overline{a}_1}{\zeta_{\rm d}} ,\\
\lambda &= \frac{1}{\zeta_{\rm d}} \left(\overline{a}_1 -\frac{\overline{b}_0}{2} \right) ,\\
\zeta &= \frac{\overline{b}_0}{\zeta_{\rm d}} .
\end{align}
The remaining portion of the dynamics in Eq.~\eqref{eq:varphiEOM2} that must be mapped to AMB+ is the noise.
In Ref.~\cite{Caballero18} the parameters $(\overline{\zeta}, \overline{u})$ appearing in the phase diagram are defined as $\overline{\zeta}=\zeta D^{1/2}K^{-3/4}$ and $\overline{u}=uDK^{-2}$ where $D$ is half the variance of the noise which must be isotropic.
Our fluctuating hydrodynamics model has an off-diagonal (i.e. anisotropic) term in the noise correlator in Eq.~\eqref{eq:noise}; however, we can safely neglect this term as we expect the bulk of the system to become isotropic near the critical point and thus expect the dominant source of noise in the critical dynamics to be isotropic and Gaussian.
When this is the case, we identify half the variance of the noise as:
\begin{equation}
    D = \frac{\ell_{\rm c} \rho_{\rm c} v_D^2}{d(d-1) U_0}.
\end{equation}
\end{subequations}

\begin{figure}[h]
	\centering
	\includegraphics[width=.45\textwidth]{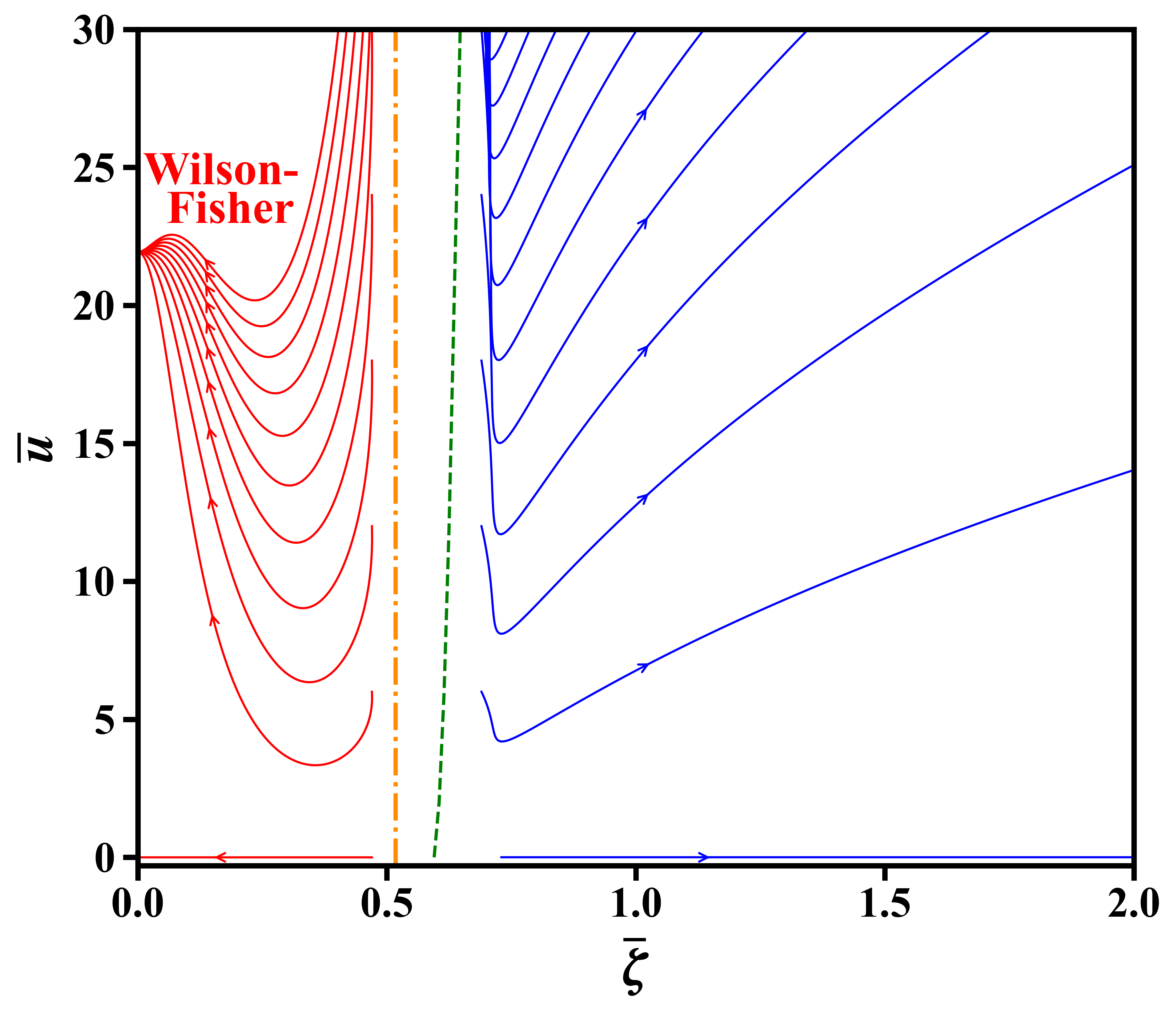}
	\caption{\protect\small{{Numerical integration of the full RG flow in Ref.~\cite{Caballero18} obtained for different initial conditions, all of them with initial parameters $\overline{\nu}=2.6$ and $\overline{\lambda}=3.25$ in $d=3$. The flow is represented in the $(\overline{\zeta}, \overline{u})$ plane. A separatrix (green dashed line) between two different regimes appears: when $\overline{\zeta}$ is small (red lines with arrow), the flow converges back to the Wilson-Fisher fixed point (we checked that $\overline{\nu}$ and $\overline{\lambda}$ are flowing to 0 in this case). On the right of the separatrix (blue lines with arrow), the flow diverges to infinity (strong coupling). The nontrivial $F_4$ fixed point resides on the separatrix, which has codimension one and thus represents a weak-to-strong coupling phase transition. The critical MIPS in our fluctuating hydrodynamics model lies on the orange dotted line (at $\overline{u} \gg 30$) which is to the left of the separatrix. The Gaussian fixed point lies at the origin.}}}
	\label{Fig:3} 
\end{figure}

We now make use of the RG equations in Ref.~\cite{Caballero18} to determine the phase diagram in $d=3$.
Figure~\ref{Fig:3} displays the resulting phase diagram.
Consistent with Ref.~\cite{Caballero18} which numerically analyzed $d=2.5$, we find a separatrix located at $\overline{\zeta} \approx 0.6$ that divides the attractive basin of the Wilson–Fisher fixed point from the strong-coupling regime.
Evaluating the equations of state for $a$ and $b$ found in Ref.~\cite{Langford24} at the critical point, we find:
\begin{equation}
    \overline{\zeta} \approx 0.524 ,
\end{equation}
which lies in the left of the separatrix.
As the flow diagram suggests, this belongs to the attractive basin of the Wilson-Fisher fixed point.
Therefore, by mapping critical MIPS to AMB+, we have established that 3D MIPS in systems of ABPs likely belongs to the Ising universality class.

A few remarks are in order.
In the above calculation, we take the critical point from the solution of Eq.~\eqref{eq:Critical} using our equations of state $(\rho_c v_D = 0.487, \ell_c/\sigma=18.199)$, which differs from the critical point determined from particle-based simulations $(\rho_c^* v_D = 0.48, \ell_c^*/\sigma=18.9)$.
Nevertheless, using $(\rho_c^* , \ell_c^*)$ gives us $\overline{\zeta} \approx 0.517$, which still lies to the left of the separatrix.
Furthermore, the nondimensionalization in Eq.~\eqref{eq:varphiEOM2} is essential for accessing the Wilson–Fisher fixed point in our model: it identifies the characteristic length scale and time scale on which the critical physics occur, below which other physics generally take over.

Interestingly, the nonlinear coupling in our model is just slightly below the separatrix located at $\overline{\zeta} \approx 0.6$.
The proximity of $\overline{\zeta}$ to the separatrix suggests that it may be possible to change microscopic details of the model (e.g., the form of the interparticle interaction potential) and drive the system across the separatrix and into the strong-coupling regime.
Despite this possibility, we are not aware of any 3D particle-based simulations that report the anticipated novel phase behavior (such as microphases~\cite{Stenhammer14,Tjhung18,Caporusso2020Motility-InducedSystem,Shi2020Self-OrganizedParticles,Semwal2024MacroSwimmers}) present in this regime.
The theoretical determination of the fixed point that the system will flow to could be made more rigorous by investigating whether higher-order terms in our model---beyond those considered in Ref.~\cite{Caballero18}---give rise to additional fixed points. 
These directions are left for future work.
Notably, recently Ref.~\cite{fejHos2025scaling} developed a nonperturbative framework to study the scaling behavior of AMB+, which confirms the existence of the strong-coupling fixed point conjectured in Ref.~\cite{Caballero18} to be associated with microphase separation.
In contrast to the perturbative RG findings, Ref.~\cite{fejHos2025scaling} found this strong-coupling fixed point does not approach the Wilson-Fisher fixed point when $d\to 2^+$, suggesting that it might not be related to bulk phase separation transition at all.

\section{Conclusions} 
We have presented extensive simulations to extract critical exponents $\nu$, $\gamma$, $\beta$, and $z$ of 3D MIPS in systems of ABPs, all of which closely match those of the 3D Ising universality class with conserved dynamics.
We developed a hydrodynamic description of the critical dynamics of 3D ABPs, which contains the nonlinear couplings that could cause a system to flow to the strong-coupling regime.
To rigorously determine if these couplings are strong enough to actually induce this strong-coupling regime, we mapped our dynamics onto the AMB+ model of Ref.~\cite{Caballero18}.
In particular, AMB+ has a parameter $\overline{\zeta}$ that controls a nonlinear coupling between gradient terms.
There is a \textit{separatrix}, where systems with values of $\overline{\zeta}$ that are beyond the separatrix will flow to the strong-coupling regime while values below the separatrix will flow to the Wilson-Fisher fixed point.
We follow the approach of Ref.~\cite{Caballero18} to precisely determine the RG flow at the one-loop level in $d=3$ and find the location of the separatrix, which we use to determine the universality class of our fluctuating hydrodynamics model.
Consistent with our numerics, we find that critical MIPS in systems of ABPs belongs to the attractive basin of the Wilson-Fisher fixed point.
Importantly, the model lies near the separatrix that separates systems that flow to the Wilson-Fisher fixed point from those that flow to the strong-coupling regime conjectured to correspond to novel phase behavior like microphases.

While our analysis shows that 3D ABPs with short-ranged repulsive interactions are generally expected to display MIPS that falls under Ising universality, changing the nature of the interparticle interactions or activity may push the system beyond the separatrix and into the strong-coupling regime.
It is our hope that the approach outlined in this work can be applied to diverse models of active matter in order to deepen our understanding of which, if any, microscopic factors can lead to departures from Ising universality.
Finally, we note that with the establishment of Ising universality, the entire phase diagram of the repulsive active Brownian hard spheres considered here resembles that of passive fluids with attractive interactions~\cite{Omar21,Evans24}: a critical point with Ising universality, a triple point, and a largely metastable liquid-gas transition.
Active Brownian particles are thus able to entirely reproduce the classical phase diagram of passive molecular systems despite breaking time-reversal symmetry. 

\section*{Data Availability}
The data that support the findings of this article are openly available~\cite{Feng2025Data}.

\begin{acknowledgments}
J.~F. acknowledges support from the UC Berkeley College of Engineering Jane Lewis Fellowship.
D.~E. acknowledges support by the U.S. Department of Defense through the National Defense Science and Engineering Graduate Fellowship Program.
This research used the Savio computational cluster resource provided by the Berkeley Research Computing program. 
\end{acknowledgments}

\end{document}

\maketitle

\pagenumbering{gobble}

\setcounter{secnumdepth}{2}
\setcounter{tocdepth}{2}
{
	\hypersetup{
		linkcolor=Black,
		citecolor=Black
	}
	\vspace{-10pt}
	\tableofcontents
}

\noindent\rule{5.0cm}{0.4pt}
	{\footnotesize
	$\\$
	{
	    {$^\dag \,$}aomar@berkeley.edu}
	}

\newpage 

\pagenumbering{arabic}

\section{Theory of Critical Fluctuating Hydrodynamics of Active Brownian Particles}
\subsection{The Fluctuating Hydrodynamics Model of ABPs \label{sec:fluc_hydro}}
The dynamics of the order parameter (or any field) can be obtained exactly from systematic coarse-graining.
These exact dynamics for a specific field are, for all but the simplest cases, coupled to additional fields. 
However, in certain limits, it may be possible to formally approximate these dynamics through closures that reduce the number of fields. 
The resulting dynamics should satisfy the known symmetries of the system and can perhaps be connected to phenomenological dynamical descriptions, such as model B. 
Beginning from the Toner-Tu model~\cite{Toner95,Toner98}, Partridge and Lee recently used symmetry, conservation considerations, and dimensionality to describe the density field dynamics of non-aligning active matter~\cite{Partridge19}.
These arguments resulted in a straightforward identification that these critical active dynamics are precisely those of model B near a critical point and are thus consistent with the Ising universality class. 

We recently derived an approximate fluctuating hydrodynamic description of overdamped active Brownian particles (ABPs) that is valid for small density flux and timescales larger than the polarization relaxation time, $\tau_R$~\cite{Langford24}.
By introducing mechanical equations of state, the dynamics for the density field are solely in terms of the local density and take the following form:
\begin{subequations}
\label{eq:fluc_hydro}
\begin{align}
    \frac{\partial \rho}{\partial t} &= - \bm{\nabla} \cdot \mathbf{J},  \\
    \mathbf{J} &= \frac{1}{\zeta_{\rm d}} \bm{\nabla} \cdot \left( \bm{\sigma}^{\rm C} + \bm{\sigma}^{\rm act} \right) + \bm{\eta}^{\rm act}, \\
    \bm{\sigma}^{\rm C} &= \left[ -p_{\rm C} (\rho)+\kappa_1(\rho) \nabla^2 \rho + \kappa_2(\rho) |\nabla \rho|^2 \right] \mathbf{I} +\kappa_3(\rho)\bm{\nabla} \rho \bm{\nabla} \rho + \kappa_4(\rho) \bm{\nabla} \bm{\nabla} \rho , \label{eq:sigma_c} \\
    \bm{\sigma}^{\rm act} &= \left[ - p_{\rm act} (\rho) + a(\rho) \nabla^2 \rho \right] \mathbf{I} +b(\rho) \bm{\nabla} \rho \bm{\nabla} \rho,\label{eq:sigma_act} \\
    \langle \bm{\eta}^{\rm act} (\mathbf{r}, t) \bm{\eta}^{\rm act} (\mathbf{r}', t') \rangle &= 2 \frac{k_B T^{\rm act}}{\zeta_{\rm d}} \left( \rho \mathbf{I} - \frac{d}{d-1} \mathbf{Q}' \right)  \delta(t-t') \delta(\mathbf{r} - \mathbf{r}'), \label{eq:noise} \\
    \mathbf{Q}' &= \frac{\rho U_0}{d p_{\rm act}} \left( -a(\rho)  \nabla^2 \rho \mathbf{I} - b(\rho) \bm{\nabla}\rho \bm{\nabla} \rho \right) .\label{eq:traceless_nematic}
\end{align}
\end{subequations}
In Eq.~\eqref{eq:fluc_hydro}, $\rho(\mathbf{r}, t)$ represents the coarse-grained density field, $\mathbf{J}$ the density flux, $\zeta_{\rm d}$ the single particle translational drag coefficient, $\bm{\sigma}^{\rm C}$ the interparticle conservative stress, $\bm{\sigma}^{\rm act}$ the active effective stress, $\bm{\eta}^{\rm act}$ athermal stochastic contribution to density flux, $\mathbf{Q}'$ the traceless nematic order density, and $d$ the dimension of the space.
Within the closures used to describe these dynamics, the traceless nematic order can be fully expressed with the density field [Eq.~\eqref{eq:traceless_nematic}].
It should also be noted that the anisotropic noise presented above is entirely athermal with the amplitude set by the active energy scale $k_B T^{\rm act} \equiv \zeta_{\rm d} U_0 \ell_0 /d(d-1)$.
The form of the noise presented in Eq.~\eqref{eq:noise} is exact for the \textit{microscopic} density and nematic field and approximate for the \textit{coarse-grained} fields of interest.

$\bm{\sigma}^{\rm C}$ [Eq.~\eqref{eq:sigma_c}] and $\bm{\sigma}^{\rm act}$ [Eq.~\eqref{eq:sigma_act}] are expanded into bulk terms ($p_{\rm C}$ and $p_{\rm act}$ respectively) and density gradients up to the second order, with $\{ \kappa_i (\rho) \}$, $a(\rho)$, and $b(\rho)$ expansion coefficients.
The gradient expansion of the conservative interaction stress in terms of the local density field necessarily restricts our analysis to ABPs with short-range interactions.
As Ref.~\cite{Omar23b} considered ABPs at high activities, the interfacial coefficients associated with the interaction stress ($\{ \kappa_i(\rho) \}$) were discarded as they fundamentally scale with the particle size ($\sigma$) while those associated with the active stress [$a(\rho)$ and $b(\rho)$] scale with the active run length ($\ell_0$).
The forms of $\{ \kappa_i(\rho) \}$ can be found for pairwise interacting systems by formally expanding the spatial kernel of the interaction stress (see, for example, Ref.~\cite{Chiu2024TheoryCoexistence}).

We now proceed to expand Eq.~\eqref{eq:fluc_hydro} about the critical point.
We introduce the order parameter $\varphi(\mathbf{r}, t)$, which is a scalar field and is a linear transform of the density field, $\varphi(\mathbf{r}, t)=\left(\rho(\mathbf{r}, t)-\rho_{\rm c}\right) v_D$, where
$\rho_{\rm c}$ is the density at the critical point.
It is worth mentioning that the following analysis can be applied to ABPs with short-range interactions, including attraction. 
However, attractive interactions are characterized by both a length and energy scale while repulsive interactions are solely characterized by a length scale in the hard particle limit. 
As our simulations are for strongly repulsive ABPs, the dimensionless group controlling our phase behavior is the ratio of the persistence length $\ell_0 \equiv U_0\tau_R$ to the particle size $\sigma$.
Defining the dynamic pressure $\mathcal{P}(\rho)=p_{\rm C}(\rho)+p_{\rm act}(\rho)$, the critical point can be determined from the following equations~\cite{Omar23b}:
\begin{equation}
\label{eq:Critical}
    \frac{\partial \mathcal{P}}{\partial \rho}\Bigg|_{\rho_{\rm c}, \ell_{\rm c}}=\frac{\partial^2 \mathcal{P}}{\partial \rho^2}\Bigg|_{\rho_{\rm c}, \ell_{\rm c}}=0 ,
\end{equation}
where the value of activity at the critical point is denoted as $\ell_{\rm c}$.
As the system approaches the critical point, Eq.~\eqref{eq:fluc_hydro} can be expanded around the critical density.
This expansion yields the dynamics for the order parameter, $\varphi(\mathbf{r}, t)$.
For the bulk term $\mathcal{P}(\rho)$ in Eq.~\eqref{eq:fluc_hydro}, we can Taylor expand about $\rho_{\rm c}$:

\begin{equation}
\label{eq:P_expansion}
   \mathcal{P}(\rho)=\mathcal{P}_{\rm c} + g_1 \varphi(\mathbf{r}, t)+g_2 \varphi^2(\mathbf{r}, t) +g_3 \varphi^3(\mathbf{r}, t) +g_4 \varphi^4(\mathbf{r}, t)+g_5 \varphi^5(\mathbf{r}, t)+ ... \ ,
\end{equation}
where each coefficient $g_n \equiv (v_D^n n!)^{-1} \partial^n \mathcal{P} / \partial \rho^n$ is evaluated at the critical point. 
As for the gradient terms in Eq.~\eqref{eq:fluc_hydro}, we note that they can be expressed as functions of the order parameter near the critical point, e.g. ${\{\kappa_i(\rho)\} \to \{\kappa_i(\varphi)\}, \bm{\nabla}\rho \to \bm{\nabla} \varphi}$.

\subsection{Standard Dimensional Analysis}
Next we conduct a standard dimensional analysis to determine the relevance of different terms with respect to the Gaussian fixed point.
As we emphasized in the main text, it would be misleading to simply discard the irrelevant terms and analyze the critical dynamics, as the irrelevant terms may also induce new fixed points.
Nevertheless, for pedagogical purposes, we will now show that model B dynamics are recovered if all irrelevant terms are discarded.
Introducing the Martin-Siggia-Rose auxiliary field $\Tilde{\varphi}$, we can write down the dynamic Janssen-De Dominicis functional of the system~\cite{Tauber14}:
\begin{subequations}
\label{eq:action}
\begin{align}
\mathcal{A}[\tilde{\varphi}, \varphi] &= \mathcal{A}_0[\tilde{\varphi}, \varphi] + \mathcal{A}_{\rm noise}[\tilde{\varphi}, \varphi], \\
\label{eq:action0}
\mathcal{A}_0[\tilde{\varphi}, \varphi] &= \int d^d r \int dt \Bigg\{ 
    \tilde{\varphi}  
        \frac{\partial \varphi}{\partial t} 
        + \tilde{\varphi} \frac{1}{\zeta_{\rm d}} \bm{\nabla} \cdot \bm{\nabla} \cdot \Bigg(
            \bigg[ 
                -g_1 \varphi - g_2 \varphi^2 - g_3 \varphi^3 
                + \kappa_1 \nabla^2 \varphi 
                + \kappa_2 |\bm{\nabla} \varphi|^2 
                 \nonumber \\
        & \quad + a \nabla^2 \varphi 
            \bigg] \mathbf{I} + \kappa_3  \bm{\nabla} \varphi \, \bm{\nabla} \varphi 
          + \kappa_4 \bm{\nabla}\bm{\nabla} \varphi 
          + b \bm{\nabla} \varphi \, \bm{\nabla} \varphi + {\rm h.o.t.} \Bigg)
    \Bigg\} , \\
\label{eq:noise_action}
\mathcal{A}_{\rm noise}[\tilde{\varphi}, \varphi] &= \int d^d r \int dt \left\{ 
    \tilde{\varphi} \bm{\nabla}\cdot  \bm{\nabla} \cdot \left[ 
        \frac{k_B T^{\rm act}}{\zeta_{\rm d}} \left( \rho \mathbf{I} - \frac{d}{d-1} \mathbf{Q}' \right) \tilde{\varphi}  
    \right]
\right\} ,
\end{align}
\end{subequations}
where ${\rm h. o. t.}$ represents higher-order terms.

Now we are prepared to derive the scaling dimensions of various quantities from Eq.~\eqref{eq:action}. 
In terms of an arbitrary momentum scale $\mu$, we have the following scaling dimensions as our starting point:
\begin{equation}
      [\mathcal{A}]=\mu^0, \quad  [r]=\mu^{-1}.
\end{equation}
We then assume that $a$ and $\kappa_1$ are kept fixed under the RG flow (i.e.,~$[a]=[\kappa_1] =\mu^0$) such that when all $g_i=0 \ (i=1,2,3,...)$, we will recover the expected Gaussian fixed point in high dimensions ($d>4$) but it becomes unstable when $d<4$~\cite{Kardar07}.
We can then obtain the scaling dimension of time:
\begin{equation}
    [t]=\mu^{-4}.
\end{equation}
Now we proceed to get the scaling dimensions of the fields.
Expanding the noise correlation in Eq.~\eqref{eq:noise_action} near the critical point, to leading order we just have Gaussian white noise, so that Eq.~\eqref{eq:noise_action} becomes $\mathcal{A}_{\rm noise}=\int d^d r \int dt \left( k_B T^{\rm act} \rho_{\rm c}/\zeta_{\rm d} \right) \Tilde{\varphi} \nabla^2 \Tilde{\varphi}$.
(The neglect of $\mathbf{Q}'$ will be discussed later.)
Therefore, with $[\mathcal{A}_{\rm noise}] = [\mathcal{A}_0] = \mu^0$ and the first term in Eq.~\eqref{eq:action0}, we conclude:
\begin{equation}
    [ \varphi(\mathbf{r}, t) ]=\mu^{-1+d/2}, \quad [ \Tilde{\varphi}(\mathbf{r}, t) ]=\mu^{1+d/2} .
\end{equation}
The dimensions of other coefficients in Eq.~\eqref{eq:action0} can now be straightforwardly determined:
\begin{align}
     [g_1]&=\mu^2, \ [g_2]=\mu^{3-d/2}, \ [g_3]=\mu^{4-d}, \ [g_4]=\mu^{5-3d/2},\ [g_5]=\mu^{6-2d}, ...  \nonumber \\
    [\kappa_1] &= [\kappa_4]=[a]=\mu^0, \ [\kappa_2]=[\kappa_3]=[b]=\mu^{1-d/2}, ...
\end{align}

In $d=3$, the gradient-expansion terms associated with the coefficients $\kappa_2$, $\kappa_3$, and $b$ are irrelevant in the RG sense, as their scaling dimensions are negative.
This also justifies omitting the higher-order terms in the expansion that were not explicitly considered in Eq.~\eqref{eq:fluc_hydro}. 
We can now revisit the contribution to the action arising from the non-diagonal noise in Eq.~\eqref{eq:noise_action}.
We first rewrite $\mathbf{Q}'$ in terms of $\varphi$:
\begin{equation}
\label{eq:Q}
    \mathbf{Q}'= A(\varphi)\nabla^2\varphi \mathbf{I}+ B(\varphi)\bm{\nabla} \varphi \bm{\nabla} \varphi ,
\end{equation}
where the specific forms of $A(\varphi)$ and $B(\varphi)$ can be found in Eq.~\eqref{eq:traceless_nematic}.
The scaling dimensions of $A(\varphi)$ and $B(\varphi)$ are found to be
\begin{equation}
    [ A]= \mu^{-1-d/2}, \ [B]=\mu^{-d},
\end{equation}
which are irrelevant in any dimension.
The \textit{diagonal} term in Eq.~\eqref{eq:noise_action} also has a dependence on $\varphi(\mathbf{r},t)$ since $\rho(\mathbf{r}, t)=\rho_{\rm c}+\varphi(\mathbf{r}, t)/v_D$.
Substituting this definition into the noise results in two terms arising from the diagonal contribution with one proportional to the constant $\rho_c$ and the other proportional to $\varphi$. 
The latter term has a scaling dimension $\mu^{1-d/2}$, which is again irrelevant in $d=3$.
The noise can thus be approximated as Gaussian and non-multiplicative in the vicinity of the critical point. 

The action near the critical point is consistent with the following dynamics in three dimensions:
\begin{subequations}
\label{eq:model_B}
\begin{align}
\label{eq:EOM_model_B}
     \frac{\partial \varphi}{\partial t}&= \frac{1}{\zeta_{\rm d}}\nabla^2\left( (g_1-a\nabla^2)\varphi + g_2 \varphi^2 + g_3 \varphi^3  + {\rm h.o.t} \right)+ \nabla \cdot \bm{\eta} , \\
     \langle \bm{\eta} (\mathbf{r}, t) \bm{\eta} (\mathbf{r}', t') \rangle &= 2 \frac{k_B T^{\rm act}}{\zeta_{\rm d}} \rho_{\rm c} \mathbf{I}   \delta(t-t') \delta(\mathbf{r} - \mathbf{r}') .
\end{align}
\end{subequations}
The presence of the term even in powers of $\varphi$ (i.e., the term proportional to $g_2$) might appear to be consistent with a first-order transition. 
For systems in equilibrium, terms of even powers on the level of the dynamics reflect the presence of odd-order terms in the free energy.
However, here it is not the case as, at criticality, we have [see Eq.~\eqref{eq:Critical}]:
\begin{equation}
    g_1=g_2=0.
\end{equation}
Equation~\eqref{eq:model_B} is thus exactly the dynamics of model B~\cite{Hohenberg77} for passive systems, but now with an athermal source of noise and activity-dependent coefficients.
We also note that $g_2$ term can be formally eliminated for all activities near the critical point by introducing a constant shift in the order parameter:
\begin{equation}
    \overline{\varphi}(\mathbf{r},t) = \varphi(\mathbf{r},t) + \frac{g_2}{3g_3} .
\end{equation}
At the critical point, the system is governed by the Wilson-Fisher fixed point, exhibiting behavior characteristic of the Ising universality class~\cite{Kardar07}.
We also note the existence of higher-order even terms like $g_4$ in Eq.~\eqref{eq:P_expansion}; nevertheless, these terms do not influence the universality class, as established by studies of asymmetric fluid criticality~\cite{Vause80,Nicoll81,Valls78,Kim03,Yarmolinsky17}.

In $d=2$, the situation is less straightforward since $[\varphi(\mathbf{r},t)]=\mu^0$, meaning the field is dimensionless. 
As a result, all coupling terms in Eq.~\eqref{eq:P_expansion} are equally relevant with $[g_i] = \mu^2$.
In the usual Ginzburg-Landau Hamiltonian for scalar order parameter with $\mathbb{Z}_2$ symmetry, the infinite sequence of coupling terms does not destabilize the Wilson-Fisher fixed point and, consequently, these terms do not alter the universality class of the system~\cite{Kardar07}.
While we can still apply similar arguments as in $d=3$ to formally eliminate the $g_2$ term through a redefinition of $\varphi$, we cannot eliminate the higher even-order terms ($g_{2j}$, $j=2, 3, ...$) which would be absent for a system with strictly $\mathbb{Z}_2$ symmetry.
However, we again reference the asymmetric fluid literature which has demonstrated that these higher-order $\mathbb{Z}_2$ symmetry violating terms do not alter the universality class~\cite{Vause80,Nicoll81,Valls78,Kim03,Yarmolinsky17}.
It may thus appear that the critical MIPS in 2D can still be described by the equation of motion [Eq.~\eqref{eq:model_B}], which continues to yield the Ising universality class in $d=2$.
However, the gradient expansion terms associated with the coefficients $\kappa_2$, $\kappa_3$, and $b$, as well as the coefficient of $\varphi(\mathbf{r}, t)$ in Eq.~\eqref{eq:noise_action}, are marginal in the RG sense.
Therefore, a more rigorous RG analysis is required to determine whether these terms are marginally relevant or marginally irrelevant, and to assess if they may destabilize the Wilson-Fisher fixed point in $d=2$.
The analysis will need to consider the non-diagonal terms and multiplicative noise which cannot be immediately discarded in two dimensions.

\subsection{Mapping to AMB+}
We now provide more details of the mapping from our fluctuating hydrodynamics model, found in Eqs.~\eqref{eq:fluc_hydro}-\eqref{eq:P_expansion} in Section~\ref{sec:fluc_hydro}, to AMB+~\cite{Tjhung18, Caballero18} discussed in the main text.
Importantly, $g_1=g_2=0$ in Eq.~\eqref{eq:P_expansion} at the critical point [see Eq.~\eqref{eq:Critical}], however we will retain $g_1$ (while dropping $g_2$) as it is generally present in AMB+ (albeit zero at the critical point).
Taylor expanding $a$ about the critical point, we have:
\begin{equation}
    \label{eq:aexpansion}
    a = a_0 + a_1 \varphi + \cdots
\end{equation}
where $a_0$ is the value of $a$ at the critical point while $a_1 \equiv d a / d \varphi|_{\varphi_{\rm c}, \ell_{\rm c}}$ where $\varphi_{\rm c} = 0$.

We now substitute Eqs.~\eqref{eq:aexpansion} and Eq.~\eqref{eq:P_expansion} into our fluctuating hydrodynamics model [Eq.~\eqref{eq:fluc_hydro}], multiply each side by $v_D$, and retain terms to $\mathcal{O}(\nabla^3 \varphi^2)$ [except for a lone cubic term of $\mathcal{O}(\nabla \varphi^3)$], resulting in the following equation of motion for $\varphi$:
\begin{multline}
    \label{eq:varphiEOM}
    \frac{\partial \varphi}{\partial t}= \frac{1}{\zeta_d} \boldsymbol{\nabla} \cdot \bigg( v_D \left( g_1 + 3 g_3 \varphi^2 \right) \boldsymbol{\nabla} \varphi - \left( a_0 + a_1 \varphi \right) \boldsymbol{\nabla} \left( \nabla^2 \varphi \right)- \left( a_1 + \frac{b_0}{v_D}  \right) \nabla^2 \varphi \boldsymbol{\nabla} \varphi \\  -  \frac{b_0}{v_D} \left(\nabla \varphi \cdot \nabla \right) \boldsymbol{\nabla} \varphi + \zeta_d \boldsymbol{\eta}^{\rm act} \bigg),
\end{multline}
where $b_0$ is the value of $b$ at the critical point.
While Eq.~\eqref{eq:varphiEOM} appears to have a form that can be mapped to AMB+ (the dynamics of which we will explicitly provide below), AMB+ is a phenomenological field theory that consequently should be dimensionless.
However, when beginning from microscopic dynamics, it is not always clear what length and time scales the dynamics of the density field should be nondimensionalized by to capture the behavior of interest.
Our dynamics have two natural length scales, the active runlength ($\ell_0$) and the particle diameter ($d_{\rm hs}$), and two natural time scales, the active reorientation time ($\tau_R$) and the relaxation time associated with translational diffusion ($\sim d_{\rm hs}^2 / U_0 \ell_0$).
We are interested in long-wavelength and long-time behavior as we are concerned with the large-scale state of matter the system will adopt.
Consequently, the dynamics should be nondimensionalized by the larger of the two length and time scales.
At the critical point, $\ell_{\rm c} / d_{\rm hs} \gg 1$ and $\tau_R /( d_{\rm hs}^2 / U_0 \ell_{\rm c}) = \ell_{\rm c}^2 / d_{\rm hs}^2 \gg 1$ and thus we nondimensionalize Eq.~\eqref{eq:varphiEOM} by the active length and time scales:
\begin{multline}
    \label{eq:varphiEOM2}
    \frac{\partial \varphi}{\partial \overline{t}}= \frac{1}{\zeta_d} \overline{\boldsymbol{\nabla}} \cdot \bigg( \left[ \overline{g}_1 + 3 \overline{g}_3 \varphi^2 \right] \overline{\boldsymbol{\nabla}} \varphi - \left( \overline{a}_0 + \overline{a}_1 \varphi \right) \overline{\boldsymbol{\nabla}} \left( \overline{\nabla}^2 \varphi \right) - \left( \overline{a}_1 + \overline{b}_0   \right) \overline{\nabla}^2 \varphi \overline{\boldsymbol{\nabla}} \varphi \\  -  \overline{b}_0  \left(\overline{\nabla} \varphi \cdot \overline{\nabla}\right) \overline{\boldsymbol{\nabla}} \varphi - \zeta_d \boldsymbol{\eta}^{\rm act} \bigg),
\end{multline}
where $\overline{t}\equiv t / \tau_{\rm R}$, $\overline{\nabla}\equiv \ell_{\rm c}\nabla$, $\overline{g}_n\equiv v_D g_n/U_0\ell_{\rm c}$, $\overline{a}_n\equiv a_n/U_0\ell_{\rm c}^3$, and $\overline{b}_0\equiv b_0/v_D U_0\ell_{\rm c}^3$.

The form of AMB+ used in Ref.~\cite{Caballero18} is:
\begin{align}
    \label{eq:AMB+eq}
    \frac{\partial \varphi}{\partial t} &= \boldsymbol{\nabla} \cdot \left(M \left[ \boldsymbol{\nabla}\left( r \varphi + u \varphi^3 - K \nabla^2 \varphi + \lambda | \nabla \varphi|^2 + \frac{\nu}{2} \nabla^2(\varphi^2)\right) + \zeta \nabla^2 \varphi \boldsymbol{\nabla} \varphi \right] - \sqrt{2 D M} \boldsymbol{\Lambda} \right) \nonumber \\ &= \boldsymbol{\nabla} \cdot \bigg(M \bigg[ \left( r + 3 u \varphi^2 \right) \boldsymbol{\nabla} \varphi - \left(K - \nu \varphi \right) \boldsymbol{\nabla} \left( \nabla^2 \varphi\right) - \left( \zeta - \nu \right) \nabla^2 \varphi \boldsymbol{\nabla} \varphi \nonumber \\ & \qquad \qquad \qquad \qquad \qquad \qquad \qquad \qquad \qquad \quad + 2 \left( \lambda + \nu \right)  \left(\nabla \varphi \cdot \nabla\right)\boldsymbol{\nabla} \varphi \bigg] - \sqrt{2 D M} \boldsymbol{\Lambda} \bigg),
\end{align}
where $r$, $u$, $K$, $\nu$, $\lambda$, $\zeta$, $M$, and $D$ are all constants and $\boldsymbol{\Lambda}$ is isotropic white noise with zero mean and unit variance.
It is clear that the deterministic part of the dynamics in Eq.~\eqref{eq:varphiEOM2} can be mapped to AMB+ [Eq.~\eqref{eq:AMB+eq}], setting $M=1$ as was done in Ref.~\cite{Caballero18}, with:
\begin{subequations}
\label{eq:mapping}
\begin{align}
r &= \frac{\overline{g}_1}{\zeta_{\rm d}}, \\
u &= \frac{\overline{g}_3}{\zeta_{\rm d}}, \\
K &= \frac{\overline{a}_0 }{\zeta_{\rm d}}, \\
\nu &= -\frac{\overline{a}_1}{\zeta_{\rm d}} ,\\
\lambda &= \frac{1}{\zeta_{\rm d}} \left(\overline{a}_1 -\frac{\overline{b}_0}{2} \right) ,\\
\zeta &= \frac{\overline{b}_0}{\zeta_{\rm d}} .
\end{align}
The remaining portion of the dynamics in Eq.~\eqref{eq:varphiEOM2} that must be mapped to AMB+ is the noise.
While $\boldsymbol{\eta}^{\rm act}$ is generally anisotropic, we expect the nematic order in the system to be negligible at the critical point and thus approximate the active noise as isotropic and Gaussian.
When this is the case, we identify half the variance of the noise as:
\begin{equation}
    D = \frac{\ell_{\rm c} \rho_{\rm c} v_D^2}{d(d-1) U_0}.
\end{equation}
\end{subequations}

In Ref.~\cite{Caballero18} the parameters $(\overline{\zeta}, \overline{u})$ appearing in the phase diagram are defined as $\overline{\zeta}=\zeta D^{1/2}K^{-3/4}$ and $\overline{u}=uDK^{-2}$.
Evaluating the equations of state for $a$ and $b$ found in Ref.~\cite{Langford24} at the critical point, we find:
\begin{equation}
    \overline{\zeta} \approx 0.524.
\end{equation}
Now that we have a mapping from our fluctuating hydrodynamics model to AMB+ model, we look to follow Ref.~\cite{Caballero18} to gain insight into the critical behavior of this system.
Although the properties of the strong-coupling fixed point identified in Ref.~\cite{Caballero18} (denoted as $F_4$ there) cannot be fully accessed perturbatively, at the time we are not aware of better approach and therefore make use of the RG equations derived in that work.

We explicitly integrate the RG equations found in Eqs.~(31)-(34) in Ref.~\cite{Caballero18}.
The full RG flow now lies in the four-dimensional parameter space $(\overline{u},\overline{\nu}, \overline{\lambda}, \overline{\zeta})$.
Making use of the symmetry of the model $(\lambda, \zeta, \nu, \varphi) \to -(\lambda, \zeta, \nu, \varphi)$~\cite{Caballero18}, we further restrict ourselves to $\lambda>0$, $u>0$ ($u<0$ is unstable).
We use the Python function \texttt{scipy.integrate.solve\_ivp} to numerically solve the ordinary differential equations governing the RG flow~\cite{Virtanen2020SciPyPython}.

For Eqs.~(31)-(34) [combined with the constants defined in Eqs.~(18)-(24)] in Ref.~\cite{Caballero18}, the only adjustable parameter in the RG equations is the ultraviolet cutoff, $q$ (defined as $\Lambda$ in Ref.~\cite{Caballero18}).
Although the universal (cutoff‐independent) features of the model should not depend on the choice of $q$, our numerical results show that certain values of $q$ can drive the RG flow toward unanticipated regions, which are most likely a manifestation of one-loop artifacts~\cite{Caballero18}. 
Accordingly, we select $q$ judiciously and benchmark our results against the results established in Ref.~\cite{Caballero18}.

Figure~3 in the main text shows our results for the full RG flow in the $(\overline{\zeta}, \overline{u})$ plane in $d=3$.
A separatrix (green dashed line) resides approximately at $\overline{\zeta} \approx 0.6$, which represents the transition from weak to strong coupling.
On the left of the separatrix, the parameters flow to the Wilson-Fisher fixed point where all three nonlinear parameters $(\overline{\lambda}, \overline{\zeta}, \overline{\nu})$ take a value of zero, and the system belongs to the Ising universality class.
On the right of the separatrix, the parameters flow to infinity and the system is conjectured to be in the strong-coupling regime.
The nontrivial fixed point $F_4$, which is specific to AMB+, resides on the separatrix.
Moreover, we find the position of the Wilson-Fisher fixed point depends on the ultraviolet cutoff $q$, while the approximate $\overline{\zeta}$ value of the location of the separatrix also depends on the specific initial values of $\overline{\lambda}$ and $\overline{\nu}$.
Nevertheless, the qualitative features of the RG flow do not change with the specific values of the parameters, and we find the system always flows to the Wilson-Fisher fixed point when $\overline{\zeta}$ lies to the left of the separatrix.

In Ref.~\cite{Caballero18} it is conjectured that the $F_4$ fixed point governs critical microphase separation while the Wilson-Fisher fixed point governs critical bulk phase separation.
Interestingly, from Fig.~3 in the main text we see that critical MIPS in our ABP model is very close to the separatrix.
This proximity suggests that a small change in the model parameters might cause the critical point of MIPS to flow to the strong-coupling regime.
This could potentially be observed in particle-based simulations, e.g. one could change the details of the interparticle potential or active noise and observe the formation of microphases.
Moreover, in Eq.~\eqref{eq:varphiEOM2} we considered terms up to order $\mathcal{O}(\nabla^3 \varphi^2)$ [except the $\overline{g}_3$ term which has order $\mathcal{O}(\nabla \varphi^3)$], just as was done in AMB+ [see Eq.~\eqref{eq:AMB+eq}].
While it remains an open question whether higher-order terms (which are generally present in the coarse-grained ABP dynamics) could induce a new strong-coupling fixed point, such RG calculations are beyond the scope of present work and are left for future study.

We note that the critical point that solves Eq.~\eqref{eq:Critical}, $(\rho_{\rm c} v_D = 0.487, \ell_{\rm c}/\sigma=18.199)$, which is what was used in the above calculation, is different from the critical point measured from simulations, $(\rho^*_{\rm c} v_D = 0.48, \ell^*_{\rm c}/\sigma=18.9)$.
If one wishes to map the dynamics of our fluctuating hydrodynamics model to AMB+ at $(\rho^*_{\rm c} v_D = 0.48, \ell^*_{\rm c}/\sigma=18.9)$, the coefficient $g_2$ (corresponding to the second derivative of $\mathcal{P}$) will generally be nonzero.
When this is the case, one can introduce a shifted order parameter $\overline{\varphi} \equiv \varphi + \Delta$, where $\Delta = g_2 / 3 g_3$, and map the result to AMB+.
This leads to $\overline{\zeta} \approx 0.517$ which is even further to left of the separatrix.
This suggests that although small changes to the equations of state can be made, one would still find the system flows to the Wilson-Fisher fixed point.

\section{Numerical Details of Finite-size Scaling Analysis}
\subsection{Brownian Dynamics Simulation Details}
The ``stiffness'' parameter $\mathcal{S} \equiv \varepsilon/(\zeta U_0 \sigma)$ determines whether hard-sphere statistics is recovered for active spheres interacting with a WCA potential.
To understand this, note that the force between a pair of particles can be expressed as $\mathbf{F}^{\rm C}_{ij}(\mathbf{x}_{ij})=-\bm{\nabla} \varepsilon u^{\rm WCA}(r;\sigma)$ where $\mathbf{x}_{ij}$ is the distance between a pair of particles and the dimensionless potential $u^{\rm WCA}$ has the following form~\cite{Weeks71}:
\begin{equation*}
u^{\rm WCA}(r;\sigma)= 
\begin{cases}
  4\left[ (\frac{\sigma}{r})^{12}-(\frac{\sigma}{r})^6 \right] +1, \ & r\le d_{\rm hs} \\
  0, & r > d_{\rm hs}
\end{cases}
\end{equation*}
where $d_{\rm hs}=2^{1/6} \sigma$. 
Selecting $\zeta U_0$ and $\sigma$ to be the units of force and length, respectively, results in a dimensionless force $\overline{\mathbf{F}}^{\rm C}_{ij}(\overline{\mathbf{x}}_{ij}; \mathcal{S})=\mathcal{S} \overline{\bm{\nabla}} u^{\rm WCA} (\overline{r})$ (where $\overline{\bm{\nabla}} \equiv \bm{\nabla}\sigma$ is the dimensionless gradient operator) that is entirely characterized by the stiffness parameter $\mathcal{S}$.
A selection of $\mathcal{S}=50$ ensures that the active force cannot generate overlaps within a pair separation distance, $d_{\rm ex}$, of $d_{\rm ex}/d_{\rm hs} \approx 0.9997$, leaving a negligible range where continuous repulsions are present.
A chosen timestep of $10^{-4} \sigma/U_0$ ensures minimal particle overlap and results in effective hard-sphere statistics in our simulation.
While there is a narrow range of particle separations in which continuous repulsions are present, hard-sphere statistics remain closely approximated.

\subsection{Theoretical Justification of the Sub-box Method}
\begin{figure*}[h]
	\centering
	\includegraphics[width=.7\textwidth]{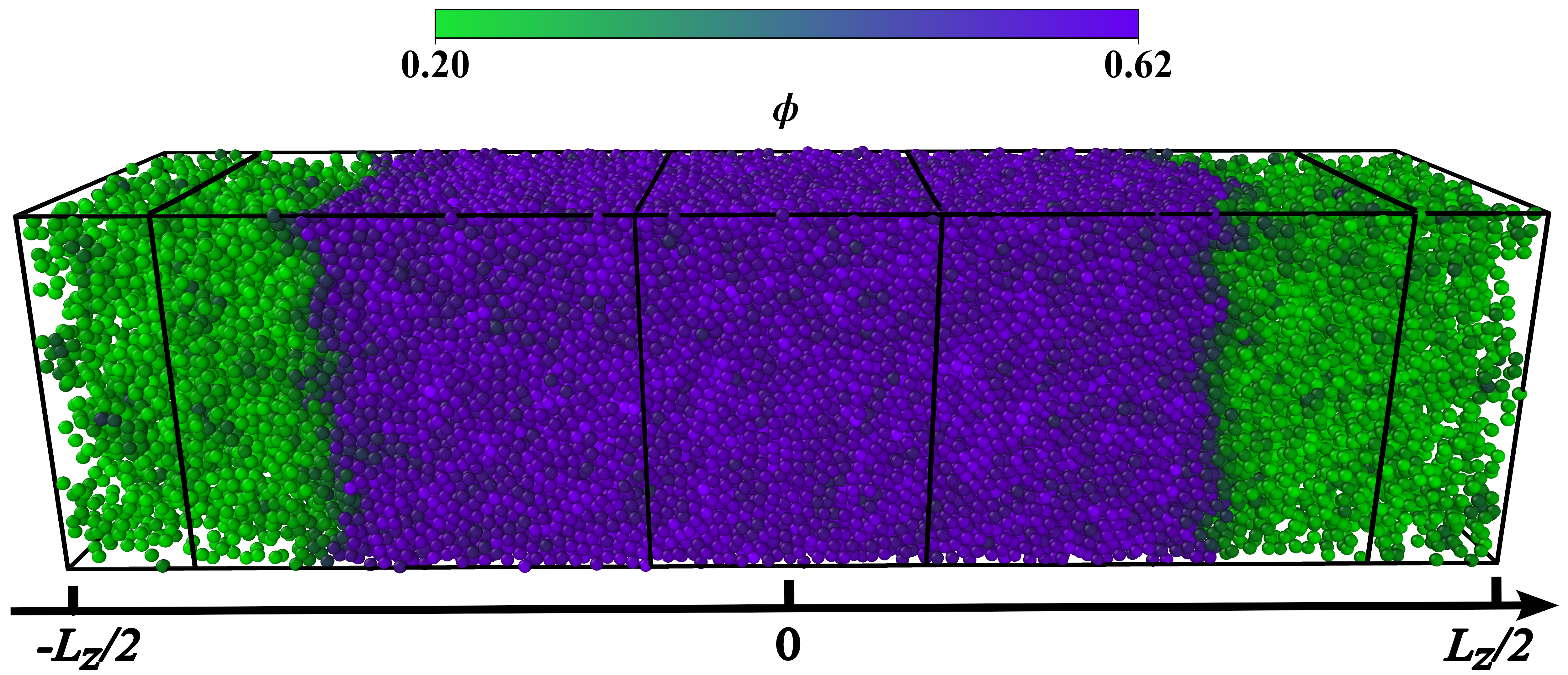}
	\caption{\protect\small{{Schematic representation of the simulation box. Simulations are done at medium packing fractions in an elongated box with an aspect ratio of 1:1:4. This results in a slab geometry where the interface is always perpendicular to the $z$-axis. Liquid and gas sub-boxes are placed at $z=0$ and $z=L_z/2$, where $z=0$ is set to be the center of mass. This snapshot uses $\ell_0/\sigma=50.0$, $\phi=0.45$, and $N=64000$.}}}
	\label{Fig:S1} 
\end{figure*}
\noindent
Finite-size scaling was initially developed for the Ising model~\cite{Binder81} and later extended to liquid-gas phase separation~\cite{Rovere90}. 
More recently, it has been successfully adapted to active systems in two dimensions~\cite{Siebert18,Partridge19,Dittrich21,Maggi21} and can be directly applied to 3D systems. 
Here, we briefly recapitulate the method and justify its usage.

In our simulation, we use an elongated simulation box with an aspect ratio of 1:1:4 to force the steady-state phase separation into a slab-like geometry. 
This setup restricts density fluctuations at the interface to the $z$-direction, allowing us to readily locate the center of the liquid phase.
Denoting the $z$-direction size of the box as $L_z$, four sub-boxes, each with a linear size of $L_z/8 (=L_x/2)$, are positioned at the liquid phase center, which coincides with the center of mass $z=0$ (see Fig.~\ref{Fig:S1} which displays a phase-separated state far from the critical point).
Four additional sub-boxes are placed at $L_z/2$ (or $-L_z/2$ due to periodic boundary condition) to sample the gas phase.
The advantage of using an elongated simulation box is that it places an upper bound of $L_x(=L_z/4)$ on the correlation length $\xi$, allowing critical density fluctuations to be sampled effectively, akin to a grand canonical ensemble~\cite{Siebert18}.

We seek to determine the distribution function of the density in each sub-box in $d$-dimension ($d=3$ in our study), $P_L(\rho)$, where $L$ is the linear dimension of the sub-box, and $\rho \equiv N_L/L^d$ is the number density.
When $L\gg \xi$, one can in principle divide each sub-box into many weakly interacting subsystems.
In the homogeneous phase, we thus expect that the distribution of density in the sub-boxes will be Gaussian~\cite{Binder81,Rovere90,Rovere93,Roman1997FluctuationsEffects,Roman98,Landau2013Statistical5,Siebert18,Partridge19}:
\begin{equation}
\label{eq:distribution}
    P_L (\rho) \propto \exp \left( -\frac{(\rho - \langle \rho \rangle)^2}{2\langle \rho \rangle} \frac{L^d}{\chi^{(L)}} \right) .
\end{equation}
In our case, we take $\langle \rho \rangle \equiv \rho_{\rm c}$ as the critical density. $\chi^{(L)} \equiv L^3 \langle \Delta \rho^2\rangle_L/\langle \rho \rangle_L $ is the susceptibility estimated within a finite box of size $L$, which is proportional to the variance over the average of the density of all sub-boxes.
As introduced in the main text, in order to locate the critical activity, we calculate the fourth-order cumulant of density fluctuations $\mathcal{B}_L\equiv \langle \Delta \rho^2\rangle^2_L/\langle \Delta \rho^4\rangle_L$, which is related with the Binder parameter $U_L$ through $U_L=1-1/3\mathcal{B}_L$~\cite{Binder81}.
Equation~\eqref{eq:distribution} (exact when $L\gg \xi$) allows us to directly calculate $\mathcal{B}_L$ in this case, and we have:
\begin{equation}
\label{eq:binder_homo}
    \mathcal{B}_L \to \frac{1}{3}, \ L\to \infty .
\end{equation}

In the phase-separated regime, the density distribution within the sub-boxes is centered around the densities of the liquid and gas phases. 
This is a result of our sub-box method, which extracts the bulk phase density distribution while excluding the interface regions.
The density distributions of liquid and gas sub-boxes are given by, respectively:
\begin{equation}
\label{eq:distribution_two}
    P_{L, {\rm liq/gas}} (\rho) \propto \exp \left( -\frac{(\rho - \rho_{\rm liq/gas})^2}{2 \rho_{\rm liq/gas}}  \frac{L^d}{\chi_{ {\rm liq/gas}}^{(L)}} \right) .
\end{equation}
The final density distribution, on which we calculate $\mathcal{B}_L$, will be an equal-weight superposition of $P_{L, {\rm gas}} (\rho)$ and $P_{L, {\rm liq}} (\rho)$ since the sub-boxes representing the gas and liquid phases have equal volumes.
From Eq.~\eqref{eq:distribution_two} we have, to leading order:
\begin{equation}
    \mathcal{B}_L \to 1, \ L\to \infty .
\end{equation}

In the vicinity of critical point, the density distribution will be non-Gaussian.
From finite-size scaling analysis, we anticipate the density distribution function take the following form~\cite{Rovere90,Rovere93}:
\begin{equation}
\label{eq:critical}
    P_L(\rho)=L^{\beta/\nu} \Tilde{P}_L((\rho-\rho_{\rm c})L^{\beta/\nu}, L^{1/\nu} \lambda) ,
\end{equation}
where we define the dimensionless distance to the critical point $\lambda \equiv (\ell_0-\ell_{\rm c})/\ell_{\rm c}$, and $\Tilde{P}_L$ is a universal function that explicitly depends on the indicated quantities.
From Eq.~\eqref{eq:critical}, it is clear that when $\rho=\rho_{\rm c}$ and $\lambda=0$, $P_L(\rho)=L^{\beta/\nu} \Tilde{P}_L(0)$, leading to the crossing of $\mathcal{B}_L$ for different system sizes.
We also have:
\begin{equation}
    \frac{\partial \mathcal{B}_L}{\partial \lambda} \propto L^{1/\nu},
\end{equation}
which is a direct implication of Eq.~\eqref{eq:critical}. 

\subsection{Error Estimation}
All quantities appearing in Fig.~1 of the main text (i.e. the Binder cumulant $\mathcal{B}$, the susceptibility $\chi$, and the phase volume fraction difference $m$) are averaged over 40000 configurations and over all sub-boxes.
Individual configurations are taken at regular time intervals separated by a duration of $20 \sigma/U_0$.
Since the relaxation time can be as long as $\sim 105 \sigma/U_0$ for the largest system even away from the critical point [see Fig.~\ref{Fig:S3}(b)], different configurations are unavoidably correlated.
In order to correctly estimate the average values and errors, we use the jackknife method as follows:
\begin{enumerate}
    \item We divide the 40000 configurations into $n$ groups, each of equal length $40000/n$;
    \item We calculate the quantity of interest using the data within each group, generating data points $\{ x_i \}$ for $i = 1, 2, \dots, n$;
    \item We compute the mean $\overline{x}_{(i)}$ of the jackknife subsample consisting of all but the $i$-th data point, and this is called the $i$-th jackknife replicate:
    \begin{equation*}
        \overline{x}_{(i)} = \frac{1}{n-1} \sum \limits_{j\neq i} x_j \ ;
    \end{equation*}
    \item These $n$ jackknife replicates $\{ \overline{x}_{(i)} \}$ give us an approximation of the distribution of the sample mean. 
    Then we have the unbiased estimation of the sample mean
    \begin{equation*}
        \overline{x} = \frac{1}{n}\sum \limits_{i=1}^n \overline{x}_{(i)} 
    \end{equation*}
    and its variance
    \begin{equation*}
        \sigma^2 = \frac{n-1}{n} \sum \limits_{i=1}^n (\overline{x}_{(i)} - \overline{x} )^2 .
    \end{equation*}
\end{enumerate}

\begin{figure*}[h]
	\centering
	\includegraphics[width=.85\textwidth]{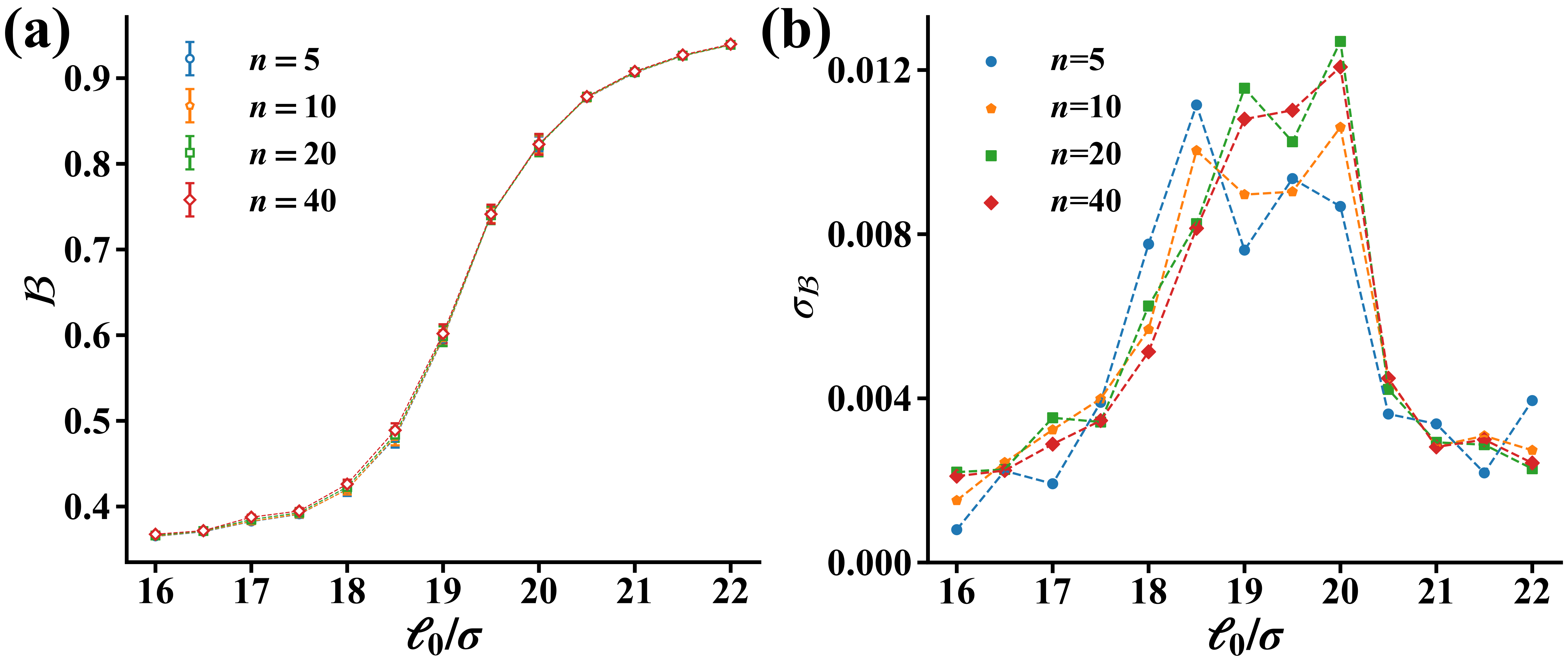}
	\caption{\protect\small{{Estimation for the (a) average value and (b) standard deviation of fourth-order cumulants $\mathcal{B}$ of $N=64000$. $n$ represents the number of groups used to divide the total data.}}}
	\label{Fig:S2} 
\end{figure*}

We here show that when $n$ increases (while maintaining a subgroup trajectory length that exceeds the relaxation time for the order parameter fluctuations), the sample mean and variance progressively converge, which validates the efficiency of our sampling method [see Fig.~\ref{Fig:S2}(a)].
As anticipated, the error is largest near the critical point [see Fig.~\ref{Fig:S2}(b)]. 
We use $n=40$ for the estimation of the average values and errors for Fig.~1 in the main text.
The errors associated with any fits (e.g., the fits displayed in the insets of Fig.~1 in the main text) are obtained through the covariance matrix obtained during our fitting procedure using \texttt{optimize.curve\_fit} in \texttt{scipy} package~\cite{Virtanen2020SciPyPython}.

\subsection{Activity and Size Dependence of Density Relaxation Time}
\begin{figure}[h]
	\centering
	\includegraphics[width=.85\textwidth]{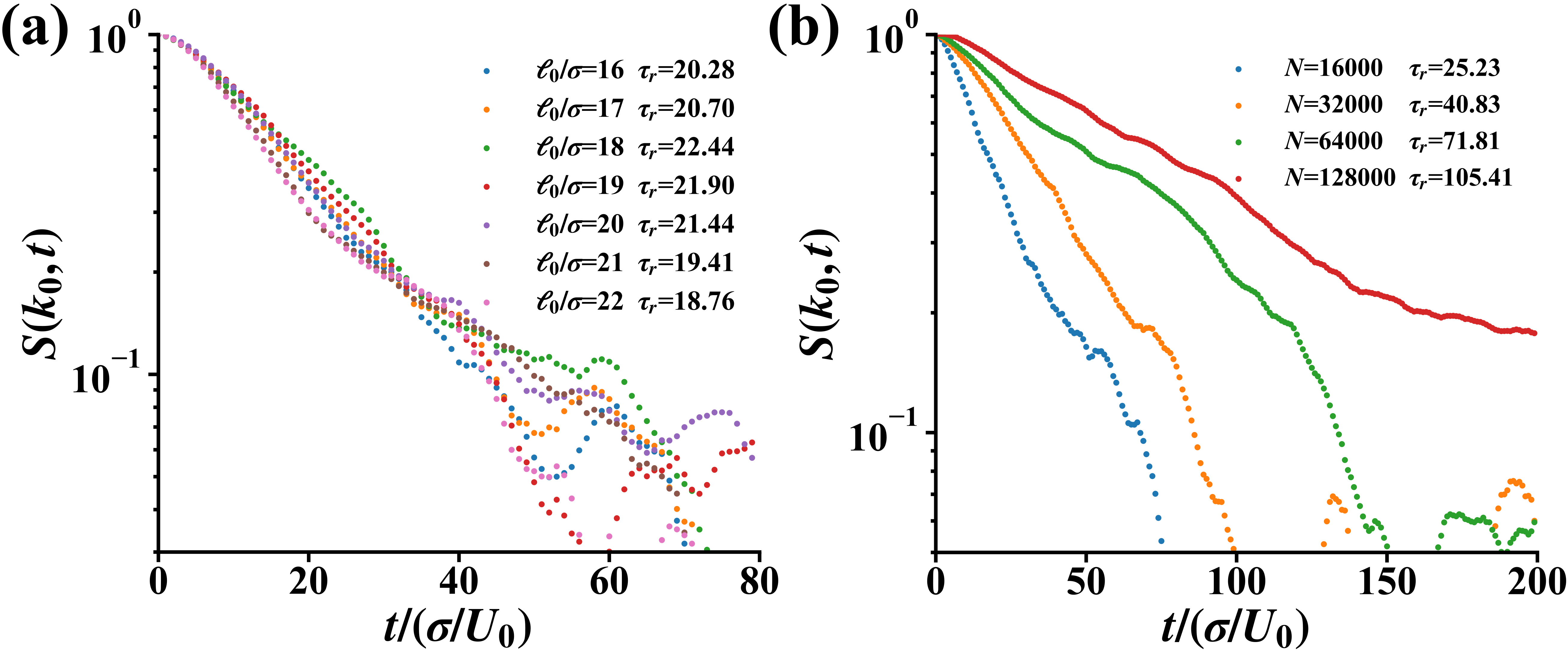}
	\caption{\protect\small{{The evolution of dynamic structure factor $S(k_0,t)$ for the system with (a) $N=16000$, $16\le \ell_0/\sigma \le 22$ and (b) $\ell_0/\sigma=18.9$, $N=(1.6, 3.2, 6.4, 12.8)\times10^4$. $\tau_r$ is the characteristic relaxation time in units of $\sigma/U_0$.}}}
	\label{Fig:S3} 
\end{figure}

To determine the simulation duration required to obtain uncorrelated configurations for our finite-size analysis, we require a measure of the relaxation time of the spatial fluctuations of our order parameter. 
One way to characterize these timescales is by calculating the dynamic structure factor, which provides insights into the spatial and temporal correlations of the density field, our order parameter.
The definition of dynamic structure factor is:
\begin{equation}
S(\mathbf{k}, t) = \frac{1}{N} \left\langle \sum_{i,j=1}^{N} \exp \left( \mathrm{i}\, \mathbf{k} \cdot \left[ \mathbf{x}_i(t) - \mathbf{x}_j(0) \right] \right) \right\rangle \equiv \frac{1}{N} \left \langle \delta \rho_{\mathbf{k}}(t) \delta \rho_{-\mathbf{k}} (0) \right \rangle ,
\label{eq:dynamic_structure_factor}
\end{equation}
which is the Fourier transform of density-density correlation function.
We fix the overall density to $\phi=0.48$ and examine the largest wavelength fluctuations [corresponding to $\mathbf{k}=\mathbf{k}_0\equiv(2\pi/L_x, 2\pi/L_y,$ $2\pi/L_z)$].
We determine $S(k_0, t)$ for $N=16000$ particles across various activity levels, as well as for different system sizes with $\ell_0/\sigma=18.9$.
The results are displayed in Fig.~\ref{Fig:S3}.

In Fig.~\ref{Fig:S3} (a) and (b), each curve is averaged over 200 and 100 independent runs, respectively.
Assuming $S(k_0,t)$ relaxes exponentially~\cite{Omar23b}, we fit its temporal evolution with an exponential decay $S(k_0,t) \propto \exp(-t/\tau_r)$.
A diffusive relaxation process would allow us to define $\tau_r=(Dk_0^2)^{-1}$ where $D$ is an activity-dependent diffusion constant.
The determination of the diffusion constant allows us to anticipate the change in the relaxation time when changing the system size through the $k_0$ dependence of $\tau_r$. 
In Fig.~\ref{Fig:S3}(a), we fit the decay of $S(k_0,t)$ until $t=50\sigma/U_0$ for a fixed system size as a function of activity with the resulting relaxation times displayed in the legend. 
Consistent with our expectation of a critical slowing down~\cite{Tauber14}, the relaxation time is maximal near the critical point. 
In Fig.~\ref{Fig:S3}(b), we report the system size dependence of the relaxation near the critical activity. 
The decay of these curves is fitted until $t=150\sigma/U_0$.
From these data, we indeed find that the size dependence of the relaxation time are consistent with a diffusive relaxation time of $\tau_r=(D k_0^2)^{-1} \propto D^{-1}N^{2/3}$ with our fitted scaling resulting in  $\tau_r \propto N^{0.663 (0.047)}$.  

Two configurations in a trajectory separated by a duration $\Delta t\gtrsim \tau_r$ should be statistically independent.
From the value of $\tau_r$ we extracted from Fig.~\ref{Fig:S3}, while for $N=16000$ our choice of $\Delta t = 20\sigma/U_0$ may be fair, for larger system sizes adjacent frames are unavoidably statistically correlated.
This motivated the previously discussed jackknife method to determine the uncertainty in our reported data.

\section{Numerical Determination of Critical Exponents}
\subsection{Determination of Critical Point}
\begin{figure*}[h]
	\centering
	\includegraphics[width=.95\textwidth]{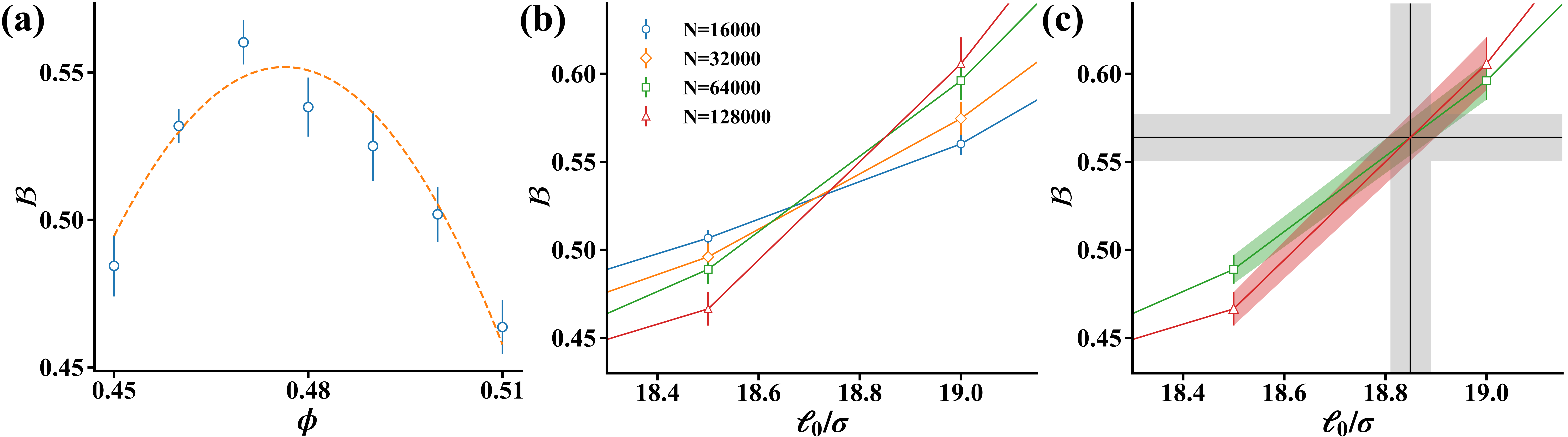}
	\caption{\protect\small{{(a) Fourth-order cumulant $\mathcal{B}$ as a function of volume fraction $\phi$. Data points are taken at fixed activity ($\ell_0/\sigma=18.9$) and number of particles 
    ($N=16000$). The dashed line is a fit to data points using second-order polynomial. (b) $\mathcal{B}$ for $\ell_0/\sigma=18.5$ and $\ell_0/\sigma=19.0$ for different system sizes, showing multiple intersection points. (c) Intersection of $\mathcal{B}$ for $N=64000$ and $N=128000$, which is used for locating the critical activity $\ell_{\rm c}/\sigma=18.850(0.048)$ and Binder cumulant $\mathcal{B}_{\rm c}=0.564(0.013)$ (black lines). The errors on $\ell_{\rm c}/\sigma$ and $\mathcal{B}_{\rm c}$ (gray areas) are obtained by propagating the $y$-error on the points near the intersection (colored areas).}}}
	\label{Fig:S4} 
\end{figure*}

\noindent
Reference~\cite{Omar21} previously estimated the critical point to be located at $\ell_c/\sigma\approx 18.8$ and $\phi_c\approx 0.483$.
These estimations were not obtained through finite-size scaling, but rather by fitting the binodal data nearest to the critical point.
Here, to estimate the critical density, we performed a density scan at fixed $\ell_0/\sigma = 18.9$ (which is close to our final estimation of critical activity $\ell_c/\sigma = 18.85$) for the smallest system investigated ($N=16000$).
The Binder cumulant $\mathcal{B}$ should exhibit a maximum at $\phi=\phi_c$ when plotted as a function of $\phi$ at the critical activity $\ell_c$~\cite{Rovere90}.
Figure~\ref{Fig:S4}(a) shows that the Binder cumulant indeed displays a maximum.
To extract $\phi_c$ we fit the data points in Fig.~\ref{Fig:S4}(a) with a second-order polynomial finding $\phi_c=0.476(0.001)$ where the fit error is reported in parenthesis.
The fit in Fig.~\ref{Fig:S4}(a) superficially appears worse than that in Ref.~\cite{Maggi21} due to the fact that we are scanning a much smaller interval of density.
We take $\phi_c=0.48$ for all the simulations discussed in the manuscript unless otherwise specified.
We assert that this will not preclude an accurate estimate of the critical activity as: 
(1) the approximate convergence of the Binder cumulant $\mathcal{B}$ curves shown in Fig.~\ref{Fig:S4}(b) for four different system sizes to a small interval;
(2) the good collapse of data points in Fig.~1(d)-(f) of the main text to a single master curve, which depends on the precise location of the critical activity;
(3) Ref.~\cite{Rovere93} found the estimation of the critical temperature (for passive liquid-gas phase separation) remarkably insensitive to the precise value of the critical density.

Figure~\ref{Fig:S4}(b) displays the data points of Binder cumulants close to the critical point for four different system sizes.
Ideally, we expect the curves to all intersect at a single point. 
However, in practice, the curves intersect over a finite but narrow range of activities.
We emphasize that this remarkably narrow range of activities is comparable to the range observed in the two-dimensional investigations of critical MIPS~\cite{Siebert18,Maggi21}, despite the increased computational cost of our finite-size three-dimensional analysis.  
From the crossings in Fig.~\ref{Fig:S4}(b) we narrow down the critical activity to lie between the interval $18.6<\ell_c/\sigma<18.9$.
In Fig.~\ref{Fig:S4}(c), we find the critical activity by locating the crossing of the cumulants for two largest system sizes $N=64000$ and $N=128000$.
We also verified that this crossing yields the best data collapse in Fig.~1(d)-(f) of the main text, compared to other crossings shown in Fig.~\ref{Fig:S4}(b).
We linearly interpolate the cumulant curves and find the critical activity $\ell_c/\sigma$ and the cumulant value $\mathcal{B}_{\rm c}$ from the intersection of the green and red lines, resulting in $\ell_c/\sigma=18.850(0.048)$ and $\mathcal{B}_{\rm c}=0.564(0.013)$.

\subsection{Determination of Static Exponent $\beta$}
\begin{figure}[h]
	\centering
	\includegraphics[width=.45\textwidth]{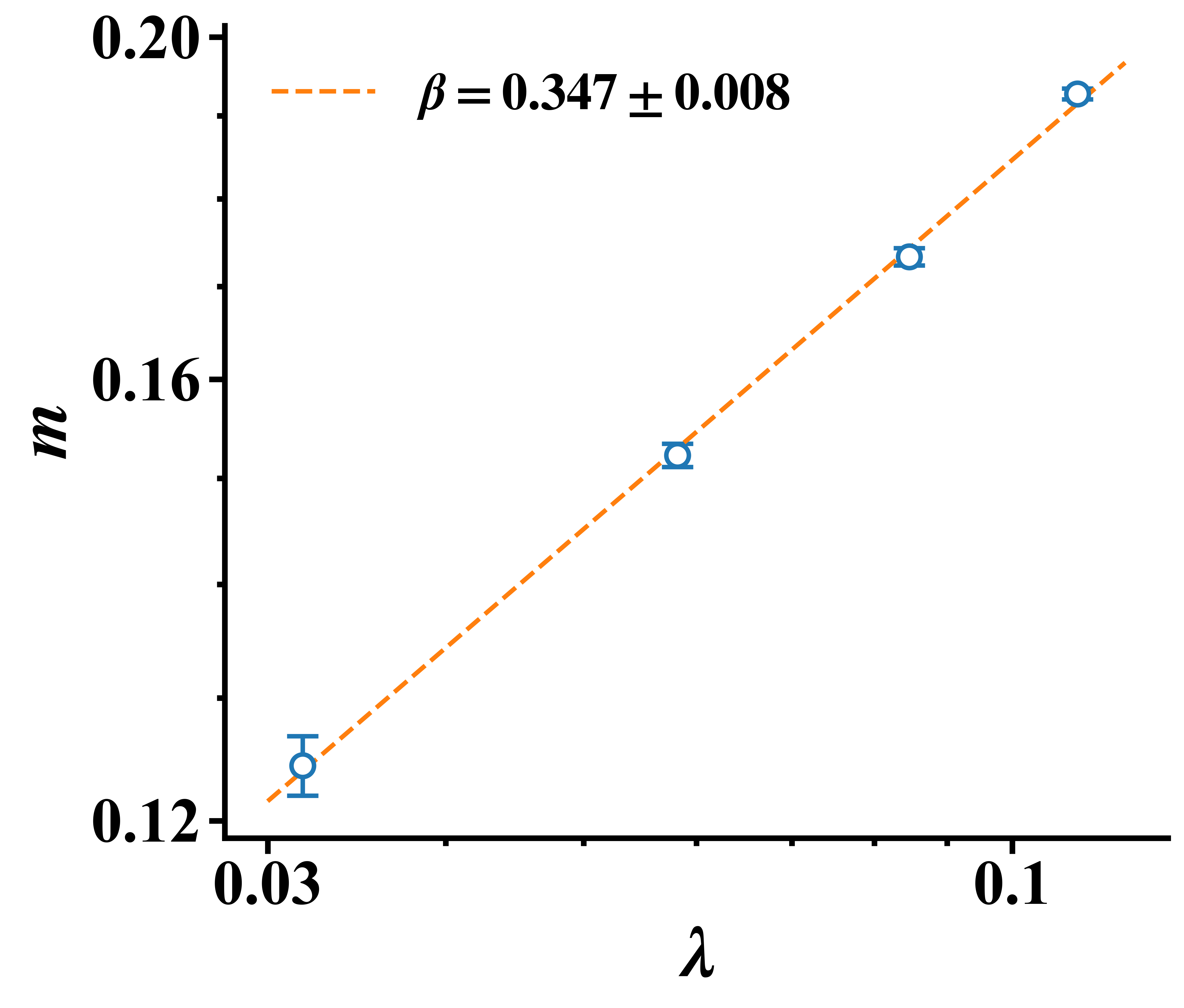}
	\caption{\protect\small{{Log-log plot of the phase volume fraction difference $m$ vs. the distance to the critical point $\lambda$ for $\ell_0/\sigma=19.5, \ 20.0 ,\ 20.5, \ 21.0$} and $N=128000$. Dashed line shows the best fit to the power law.}}
	\label{Fig:S5} 
\end{figure}

\noindent
Defining the reduced activity as $\lambda \equiv (\ell_0-\ell_c)/\ell_c$ and volume fraction difference between liquid and gas phases as $m \equiv \phi_{\rm liq}-\phi_{\rm gas}$, we have $m\propto \lambda^{\beta} (\lambda>0)$.
Fitting the difference in coexisting liquid and gas volume fractions close to the critical point is an alternative way to determine the critical exponent $\beta$~\cite{Siebert18,Dittrich21,Omar21}.
We have done so for all four sizes in our simulations, yielding, from our smallest system size to our largest,  $\beta=0.304(0.028)$, $0.343(0.006)$, $0.344(0.005)$, $0.347(0.008)$ (see Fig.~\ref{Fig:S5} for the largest system size), which are close to the 3D Ising value $\beta=0.3264...$.
Our finding thus stands in contrast to previous studies of 2D MIPS, where off-lattice~\cite{Siebert18} and on-lattice~\cite{Dittrich21} one both concluded that $\beta$ is distinctly different from the 2D Ising value.
Note that although Ref.~\cite{Siebert18} employed a different definition of $\lambda$, this does not qualitatively affect our conclusion as, in the immediate vicinity of the critical point, the leading order contribution to $m$ is insensitive to the precise definition of the rescaled activity (so long as they are zero at the critical point). 
For instance, changing the definition of reduced activity to $\lambda'=(\ell_0^{-1}-\ell_c^{-1})/\ell_c^{-1}$ and assuming $m\propto \lambda'^{\beta'} (\lambda'>0)$, we have $\beta'=0.369(0.011)$ for our largest system size, which remains consistent with the 3D Ising value.

\subsection{Determination of Static Exponent $\eta$ from the Static Structure Factor}
\begin{figure}[h]
	\centering
	\includegraphics[width=.85\textwidth]{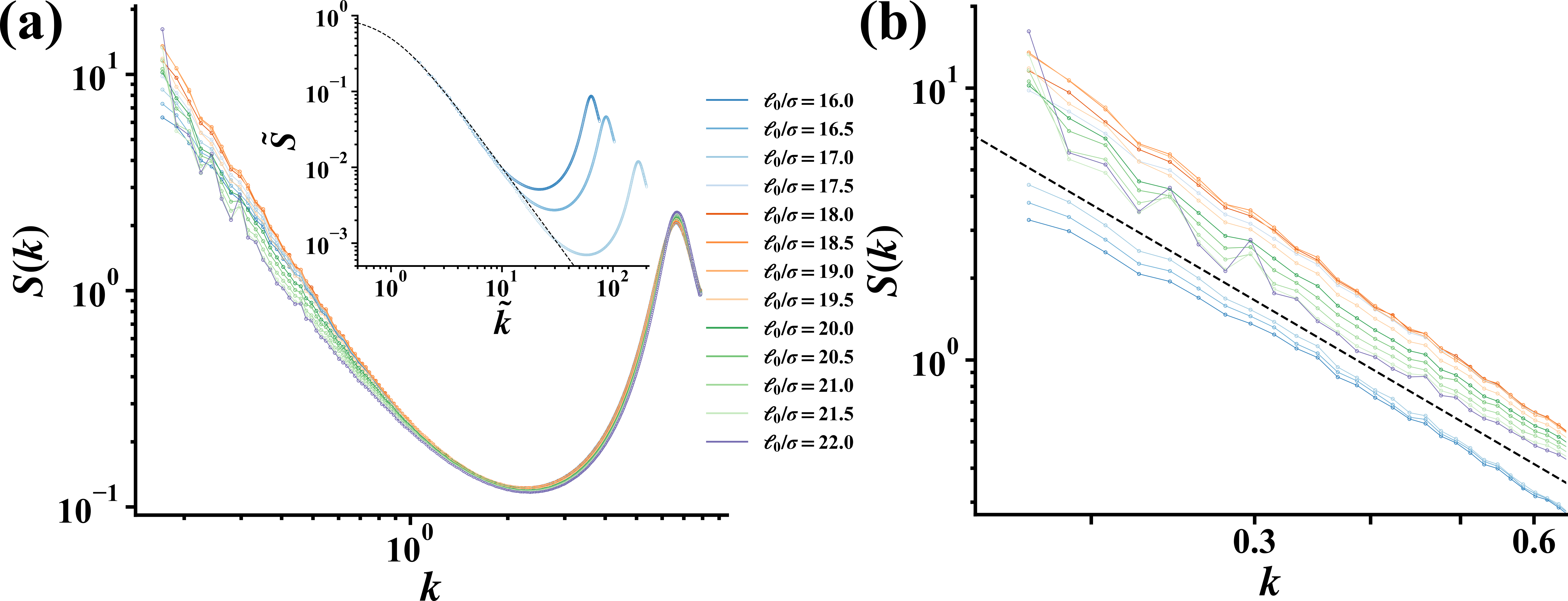}
	\caption{\protect\small{{(a) Static structure factor $S(k)$ for activities $16.0\le\ell_0/\sigma\le 22.0$ at $\phi=0.48$, $N=128000$. Inset: Rescaled structure factor $\Tilde{S}(\Tilde{k})$ (see text) for $\ell_0/\sigma=16.0, 16.5, 17.0$. (b) A magnified view of $S(k)$ at small $k$, with a vertical offset for $\ell_0/\sigma=16.0, 16.5, 17.0$. The dashed line indicates the scaling $S(k) \propto k^{-2}$.}}}
	\label{Fig:S6} 
\end{figure}

The spatial correlations of density fluctuations can be quantified through the static structure factor, which is defined as:
\begin{equation}
S(\mathbf{k}) = \frac{1}{N} \left\langle \left| \sum_{j=1}^{N} \exp \left( \mathrm{i}\, \mathbf{k} \cdot \mathbf{x}_j \right) \right|^2\right\rangle \equiv \frac{1}{N} \left \langle \delta \rho_{\mathbf{k}} \delta \rho_{-\mathbf{k}} \right \rangle .
\label{eq:static_structure_factor}
\end{equation}
We calculate the static structure factor for our largest system ($N=128000$) at the critical density ($\phi_{\rm c}=0.48$) for different activities, with the results displayed in Fig.~\ref{Fig:S6}(a).
In the homogeneous phase away from the critical point, $S(k)$ is expected to follow the form of a Lorentzian at small $k$ (the so-called Ornstein-Zernike form)~\cite{Fisher64,Fily12,Siebert18}:
\begin{equation}
\label{eq:ornstein-sernike}
    S(k)=\frac{S_0}{1+(\xi k)^2},
\end{equation}
where $\xi$ is the correlation length of the system, and $S_0$ is a free fit parameter.
At criticality $\xi$ diverges and the low-$k$ part of $S(k)$ should be well described by~\cite{Brezin74,Siebert18,Maggi21}:
\begin{equation}
\label{eq:scaling_structure}
    S(k) \propto k^{-2+\eta}
\end{equation}
for  $2\pi/\xi \ll k \ll 2\pi /a$ where $a$ is the typical distance between nearest neighbor particles at the chosen density.
Here, $\eta$ is a static critical exponent.
Reference~\cite{Siebert18} found an unphysical value for this exponent ($\eta<0$) by fitting the low-$k$ part of $S(k)$ with the authors concluding that finite-size effects prevent a reasonable extraction of the power-law behavior to the system they studied.

We fit the low-$k$ part of $S(k)$ to the form of Eq.~\eqref{eq:ornstein-sernike} and find that only the fittings of $\ell_0/\sigma=16.0,16.5,17.0$ give us a sensible correlation length $\xi$ which is smaller than system size $L_x$ (recall $L_x=L_y=L_z/4$).
In the inset of Fig.~\ref{Fig:S6}(a) we show a collapse of the rescaled structure factor $\Tilde{S}=\Tilde{S}(\Tilde{k})=S/S_0$ with $\Tilde{k}=k\xi$ and $S_0$ and $\xi$ obtained from the fits.
For $\ell_0/\sigma \ge 17.5$, the fittings give us an unphysical $\xi$ which is much larger than the system size and does not lead to the same collapse of the scaled structure factor.
This is consistent with the fact that, when fitting the low-$k$ part of $S(k)$ to Eq.~\eqref{eq:scaling_structure}, for $\ell_0/\sigma \ge 17.5$ we obtain $\eta<0$ [see Fig.~\ref{Fig:S6}(b)].

Although the exact origins for the unphysical correlation lengths and $\eta$ obtained from this procedure are unclear, we conjecture that interface created when the system phase separates to be one possible origin.
Notably, the problematic fits of $\eta$ begin to occur at activities below the  critical point $\ell_0/\sigma=18.9$.
These near-critical regions, while not yet phase-separated, exhibit near-critical fluctuations which contain interfaces between transient liquid and gas-like domains. 
We speculate that these fluctuating interfaces~\cite{Family90,Lopez97,Lassig98} could perhaps contribute to the stronger scaling of $S(k)$ with wavevector than anticipated in the absence of interfacial effects. 
Further investigation is required to resolve the discrepancy between the critical exponent obtained from the low-$k$ scaling of the static structure factor and that obtained from our sub-box procedure.

\subsection{Determination of Dynamic Exponent $z$ from Coarsening Dynamics}
To determine the dynamic exponent, we use a cubic simulation box with size $L$ instead of rectangular geometry employed in the finite-size analysis and a smaller timestep of $10^{-5} \sigma/U_0$.
The correlation length $\xi(t)$ is defined as the inverse of the first moment of the static structure factor $S(k,t)$~\cite{Stenhammar2013ContinuumParticles,Stenhammer14,Wittkowski14,Mandal19}:
\begin{equation}
\label{seq:correlation}
    \xi (t) = 2\pi \left[ \frac{\int_{k_{\rm min}}^{k_{\rm cut}} kS(k,t)dk}{\int_{k_{\rm min}}^{k_{\rm cut}} S(k,t)dk} \right]^{-1},
\end{equation}
where $k_{\rm min}$ is the smallest wave vector allowed by the finite box, $k_{\rm min}=2\pi/L$.
The upper cutoff, $k_{\rm cut}$, is chosen as the position of the first minimum in $S(k,t)$.

Figure~2 in the main text presents the dynamical scaling result for $N=10^6$ after quenching from the homogeneous phase (generated from a steady state configuration with $\ell_0/\sigma =5.0$) to conditions within the two-phase region ($\ell_0/\sigma =30.0$, $\lambda\approx 0.59$, green triangles) and the critical point ($\ell_0/\sigma =18.9$, $\lambda = 0$, purple circles).
As in many passive systems, quenches deep into the two-phase region of MIPS also lead to a characteristic domain size growing as $\xi(t) \propto t^{\alpha}$ where $\alpha = 1/3$~\cite{Puri2004KineticsTransitions,Bray2002TheoryKinetics,Stenhammer14}.
Here, our simulation reproduced the classical growth law for quenches at activities $\lambda > 0$ with a scaling of $\alpha = 0.3349(0.0015)$ for $150\le t/(\sigma/U_0)\le 1000$ in Fig.~2.
From Fig.~2 we see that the process of domain size growing is separated into three stages: one superdiffusive regime (which is believed to originate from ballistic transport~\cite{Stenhammer14}), one classical $t^{1/3}$ regime, followed by one regime where $\xi(t)$ saturates to its steady-state value which is close to $L$.
We emphasize that finite-size effects would be one possible obstacle for the observation of a non-standard exponent: Ref.~\cite{Stenhammer14} highlighted out that the box length $L$ needs to greatly exceed the persistence length fixed by $\ell_0$.
Although Ref.\cite{Stenhammer14} uses $N\approx 4\times 10^7$ to observe the  diffusive $t^{1/3}$ law, we have successfully reproduced it with far fewer particles ($N=10^6$).
This can be attributed to our shallower quench ($\ell_0/\sigma=30$ vs. $\ell_0/\sigma \approx 50$ in Ref.~\cite{Stenhammer14}), which requires a much smaller $L$.

Quenching to the critical point results in a distinct growth process. 
The three distinct regimes for the growth of the correlation length at criticality include a short-time subdiffsive regime, followed by a ``critical regime'', and finally a regime where $\xi(t)$ saturates to its steady-state value.
The critical regime takes place within $200\le t/(\sigma/U_0)\le 1200$ in Fig.~2.
During this time, the correlation length's power law exponent is fitted to be $\alpha = 0.2529(0.0042)$ which translates to a dynamic exponent of $z=3.955(0.066)$, close to the expectation of model B dynamics $z=4-\eta=3.9637...$.
It is likely that the duration of this critical regime could be extended if larger system sizes are employed as this would delay the onset of the final ``plateau'' regime.  
Interestingly, it appears that the steady-state (terminal) value of $\xi(t)$ for the deeper quench is appreciably larger than that for the critical regime, which appears contradictory with critical scaling $\xi \sim |\lambda|^{-\nu}$.
However, this is due to the nature of liquid-gas phase separation with a fixed overall density.
The constraint of fixed density ensures that for deep quenches, we will observe the coexistence of liquid and gas domains.
While the characteristic correlation length within each domain is finite, our definition of correlation length takes into account the system in its entirety, resulting in an apparently large correlation length. 
For equilibrium one-component phase separation, use of a constant pressure or chemical potential ensemble could ensure that only a single phase is present as the system density is unconstrained in these ensembles. 

\clearpage

\addcontentsline{toc}{section}{References}
\bibliographystyle{bibStyle}